\documentclass[fleqn,usenatbib]{mnras}

\usepackage{newtxtext,newtxmath}

\usepackage[T1]{fontenc}
\usepackage{ae,aecompl}

\usepackage{graphicx}   
\usepackage{amsmath}    
\usepackage{amssymb}    
\newcommand{\Cloudy}{\textsc{Cloudy}}

\font\manual=manfnt at 7pt \def\dbend{\hbox{\raise0.9ex\hbox{\manual\char127\hspace{0.6em}}}}

\providecommand{\e}[1]{\ensuremath{\times 10^{#1}}}

\newcommand\Ion[2]{\ensuremath{\mathrm{#1\,\scriptstyle #2}}}

\newcounter{INTERNALionstage}
\providecommand{\ion}[2]{
  \setcounter{INTERNALionstage}{#2}%
  \Ion{#1}{\Roman{INTERNALionstage}}}

\def\gtsim{\mathrel{\hbox{\rlap{\hbox{\lower4pt\hbox{$\sim$}}}\hbox{$>$}}}}
\def\lesssim{\mathrel{\hbox{\rlap{\hbox{\lower4pt\hbox{$\sim$}}}\hbox{$<$}}}}

\def\cm{{\rm\thinspace cm}}

\def\erg{{\rm\thinspace erg}}

\def\K{{\rm\thinspace K}}

\def\km{{\rm\thinspace km}}

\def\micron{\hbox{$\mu$m}}
\def\Mpc{{\rm\thinspace Mpc}}

\def\ps{{\rm\thinspace s^{-1}}}
\def\pcc{{\rm\thinspace cm^{-3}}}

\def\s{{\rm\thinspace s}}
\def\sr{{\rm\thinspace sr}}

\def\ergps{\mbox{$\erg\s^{-1}\,$}}

\def\kmps{\mbox{$\km\ps\,$}}
\def\kmpspMpc{\mbox{$\kmps\Mpc^{-1}\,$}}

\def\ps{\mbox{$\s^{-1}\,$}}
\def\pscm{\mbox{$\cm^{-2}\,$}}

\def\pscm{\mbox{$\cm^{-2}\,$}}

\def\ciii{\mbox{{\rm C~{\sc iii}}}}
\def\civ{\mbox{{\rm C~{\sc iv}}}}

\def\mgii{\mbox{{\rm Mg~{\sc ii}}}}

\def\feii{\mbox{{\rm Fe~{\sc ii}}}}
\def\feiii{\mbox{{\rm Fe~{\sc iii}}}}

\def\h0{\mbox{{\rm H}$^0$}}
\DeclareMathAlphabet{\vib}{OML}{cmm}{m}{it}

\newcommand{\siiii}{\ion{Si}{iii}}
\newcommand{\kms}{km\,s$^{-1}$}

\title[Fe III emission in quasars]{Fe III emission in quasars: evidence for a dense turbulent medium}

\author[M. J. Temple et al.]{Matthew J. Temple,$^{1}$\thanks{E-mail: mtemple@ast.cam.ac.uk}
Gary J. Ferland,$^{2}$
Amy L. Rankine,$^{1}$
Paul C. Hewett,$^{1}$
\newauthor
N. R.~Badnell,$^{3}$
Connor P. Ballance,$^{4}$
Giulio Del Zanna,$^{5}$ and
Roger P. Dufresne$^{5}$
\\
$^{1}$Institute of Astronomy, University of Cambridge, Madingley Road, Cambridge CB3 0HA, UK\\
$^{2}$Department of Physics and Astronomy, The University of Kentucky, Lexington, KY 40506, USA\\
$^{3}$Department of Physics, University of Strathclyde, Glasgow, G4 0NG, UK\\
$^{4}$Centre of Theoretical Atomic, Molecular and Optical Physics, Queen's University Belfast, Belfast BT7 1NN, UK\\
$^{5}$Department of Applied Mathematics and Theoretical Physics, University of Cambridge, Wilberforce Road, Cambridge CB3 0WA, UK\\
}

\date{Accepted 2020 June 12. Received 2020 June 12; in original form 2020 February 12}

\pubyear{2020}

\begin{document}
\label{firstpage}
\pagerange{\pageref{firstpage}--\pageref{lastpage}}
\maketitle

%
\begin{abstract}
Recent improvements to atomic  energy-level data allow, for the first time, accurate predictions to be made for the \ion{Fe}{iii}\ line emission strengths in the  spectra of luminous, $L_\text{bol}\simeq10^{46}-10^{48}\ergps$, Active Galactic Nuclei. The \ion{Fe}{iii}\ emitting gas must be primarily photoionized, consistent with observations of line reverberation.  We use {\sc Cloudy} models exploring a wide range of parameter space, together with $\simeq$26,000 rest-frame ultraviolet spectra from the Sloan Digital Sky Survey, to constrain the physical conditions of the line emitting gas. 
The observed \ion{Fe}{iii} emission is best accounted for by dense ($n_H\simeq 10^{14}$\,cm$^{-3}$) gas which is microturbulent, leading to smaller line optical depths and fluorescent excitation.
Such high density gas appears to be present in the central regions of the majority of luminous quasars.
Using our favoured model, we present theoretical predictions for the relative strengths of the \ion{Fe}{iii} UV34 $\lambda\lambda$1895,1914,1926 multiplet. 
This multiplet is blended with the \ion{Si}{iii}] $\lambda$1892 and \ion{C}{iii}] $\lambda$1909 emission lines and an accurate subtraction of UV34 is essential when using these lines to infer information about the physics of the broad line region in quasars.

\end{abstract}




\section{Introduction}
\label{sec:intro}

Iron lines have long been recognised as an important component in the spectra of active galactic nuclei (AGN) and quasars.
For example, emission from the \feii\ ion has been identified as one of the major sources of cooling in the broad line region \citep[BLR;][]{Wills1985} and an important contributor to the observed population variance within optical quasar spectra \citep[the so-called `eigenvector 1';][]{BG92}. 
There is now an extensive literature investigating the properties of the low-ionization (16.2\,eV) \feii\ emission in quasars and AGN. Recent results of reverberation-mapping campaigns indicate that the \feii\ emission originates in gas at distances comparable to, or larger than, the gas responsible for much of the hydrogen Balmer emission in both low and high luminosity AGN \citep[e.g.][]{Hu15,Zhang19}.

Empirical iron templates such as those provided by \citet{2001ApJS..134....1V} have also identified \feiii\ (ionization potential 30.6\,eV) as a significant source of emission, which  needs to be accounted for when modelling emission lines such as the \ciii] $\lambda1909$ blend. However, until recently the electronic energy structure of the Fe$^{2+}$ ion was poorly known and so theoretical predictions for the \feiii\ line ratios and strengths were not accurate.
Within the past few years, work by \citet{2014ApJ...785...99B} has produced improved atomic data for \feiii\ which, for the first time, allows predictions to be made for the full emission-line spectrum of this ion \citep{Laha17}.

Previous observational studies of \feiii\ have focused on the complex of lines at $\sim$2075\,\AA, which is relatively isolated and measurements are thus relatively straightforward.
\citet{2018ApJ...859...50F} find evidence that the complex of \feiii\ lines at $\lambda\lambda2039$-2113 is strongly microlensed in a sample of 11  gravitationally lensed high luminosity quasars, suggesting the line emitting region is no more than a few light days across.
At lower luminosity, \citet{2018ApJ...862..104M, 2019ApJ...880...96M}  suggest that the \feiii\ in NGC~5548 reverberates with a time scale of around three days, which is  shorter than the predicted time of 10-20 days estimated for \feii\ reverberation in the same object \citep{Hu15}.
Thus, over an extended range of luminosities, investigation of \feiii\ emission can probe the conditions of gas in quasars and AGN  closer to the central ionizing source than is the case for \feii\ emission. 

The structure of this paper is as follows.
In Section~\ref{sec:theory} we discuss the theory of the Fe$^{2+}$ ion and constrain its excitation mechanism using existing high signal-to-noise ratio composite spectra in the rest-frame ultraviolet. 
Spectra from the fourteenth data release of the Sloan Digital Sky Survey \citep[SDSS DR14Q;][]{DR14Q} provide the basis for an investigation of the statistical properties of \feiii\ emission in the population of high luminosity, $\log_{10}(L_\text{bol}/\ergps)\simeq 46.5$, quasars with redshifts $1.2<z<2.3$. 
In Section~\ref{sec:observations} we outline the selection of a sample of such objects from the DR14Q catalogue to investigate the \feiii\ emission across the quasar population.
Initial investigation of the sample is used to demonstrate the presence of \feiii\ emission with significant equivalent width at wavelengths $\simeq$1850-2150\,\AA. The observation is used to motivate additional theoretical investigation in Section~\ref{sec:models}, where we present further results of \Cloudy\ models.
The particular focus is to place constraints on the temperature, density and turbulence of the \feiii\ emitting gas. In Section~\ref{sec:line_fits} we then check the consistency of our preferred model with a more involved consideration of the \feiii\ emission properties using the full quasar sample.
The implications for our understanding of quasar broad-line regions form the basis for the discussion in Section \ref{sec:discuss}. 
The paper concludes with a short summary of the main results in Section \ref{sec:conclude}.

We assume a flat $\Lambda$CDM cosmology throughout this work, with $\Omega_m=0.3$, $\Omega_{\Lambda}=0.7$, and $\textrm{H}_0=70 \kmpspMpc$. 
All emission lines are identified with their wavelengths in vacuum in units of \AA ngstr\"{o}ms.

\section{Formation of \feiii \ lines in AGN}
\label{sec:theory}

\begin{figure}
\begin{center}
\includegraphics[clip=on,width=\columnwidth,height=0.8
\textheight,keepaspectratio]{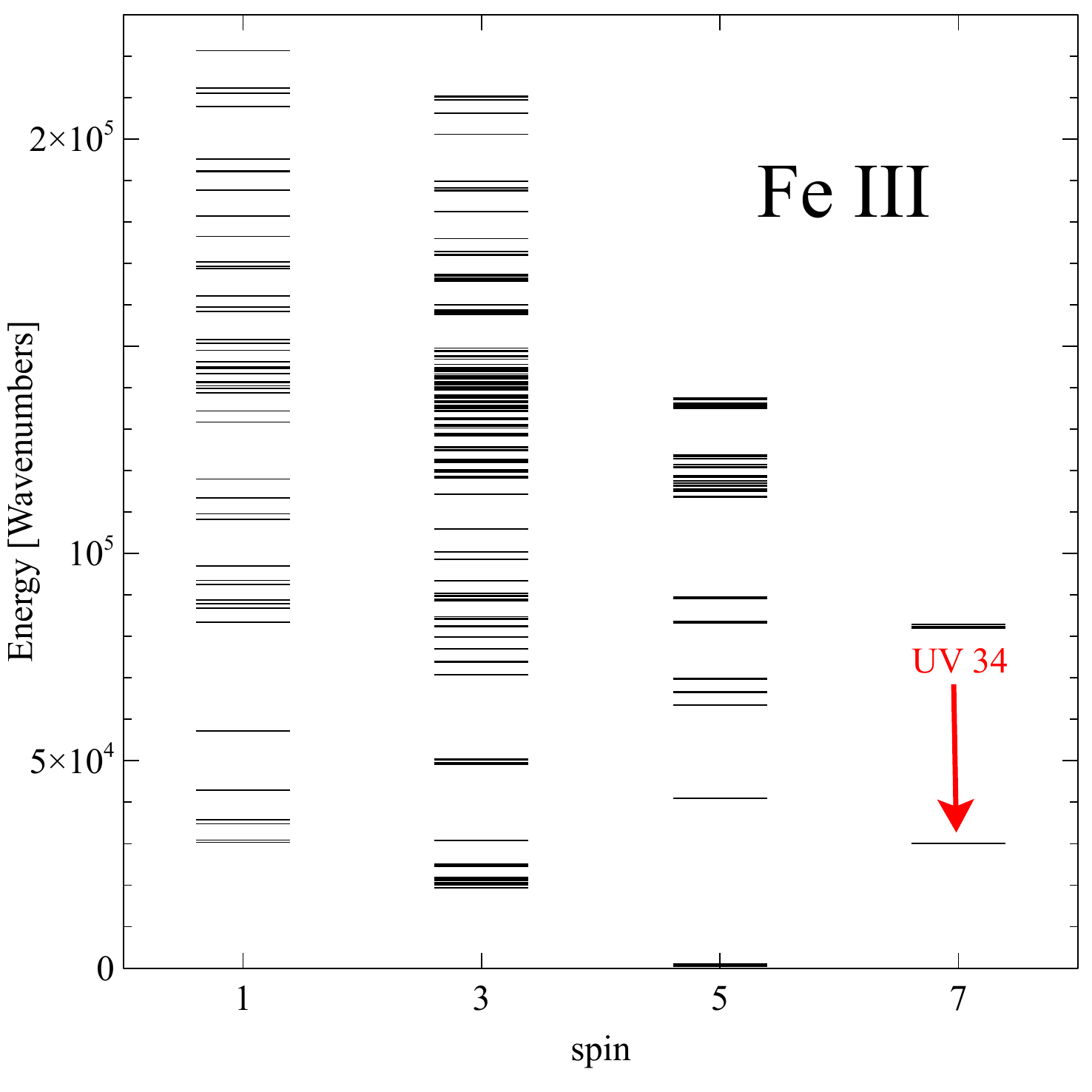}
\end{center}
\caption{The \feiii\ model now implemented in Cloudy. 
The UV34 transition studied in this paper is indicated.}
\label{fig:Fe3energies}
\end{figure}

Spectral calculations are performed using version 17.02 of \Cloudy,
last described by \citet{2017RMxAA..53..385F}.
Figure~\ref{fig:Fe3energies} shows the \feiii\ model now implemented in \Cloudy.
Data are largely from \citet{2014ApJ...785...99B} with experimental
energies from the National Institute of Standards and Technology (NIST) atomic spectra database \citep{NIST_ASD} adopted where possible.
Previous work on the \feiii \ ion is summarized by \citet{2014ApJ...785...99B}.

This paper centres on \feiii\ UV34 \citep[as defined by][]{Moore52} at
$\lambda\lambda 1895,1914,1926$ resulting from the septet transition indicated in
Fig.~\ref{fig:Fe3energies} and also the strongest \feiii\  multiplet
indicated in the \citet{2001ApJS..134....1V} template.
According to NIST the next higher septet is $^7D$ at  147\,000 wavenumbers,
a level not included in  \citet{2014ApJ...785...99B}.
Little is known about the emission properties of \feiii \
from the dense gas found near the centres of AGN and we begin with some fundamental considerations.

Lines can form via two processes, photoionization or collisional ionization.
The key difference is in the gas kinetic temperature,
with photoionized gas having a temperature set by energy balance,
generally around $10^4\K$
\citep{2006agna.book.....O}, while the gas kinetic temperature is near
the ionization temperature of the ion in collisional equilibrium,
around $10^{4.5}\K$ for Fe$^{2+}$
\citep{Lykins.M13Radiative-cooling-in-collisionally-ionized}.

\begin{figure}
\begin{center}
\includegraphics[clip=on,width=\columnwidth,height=0.8
\textheight,keepaspectratio]{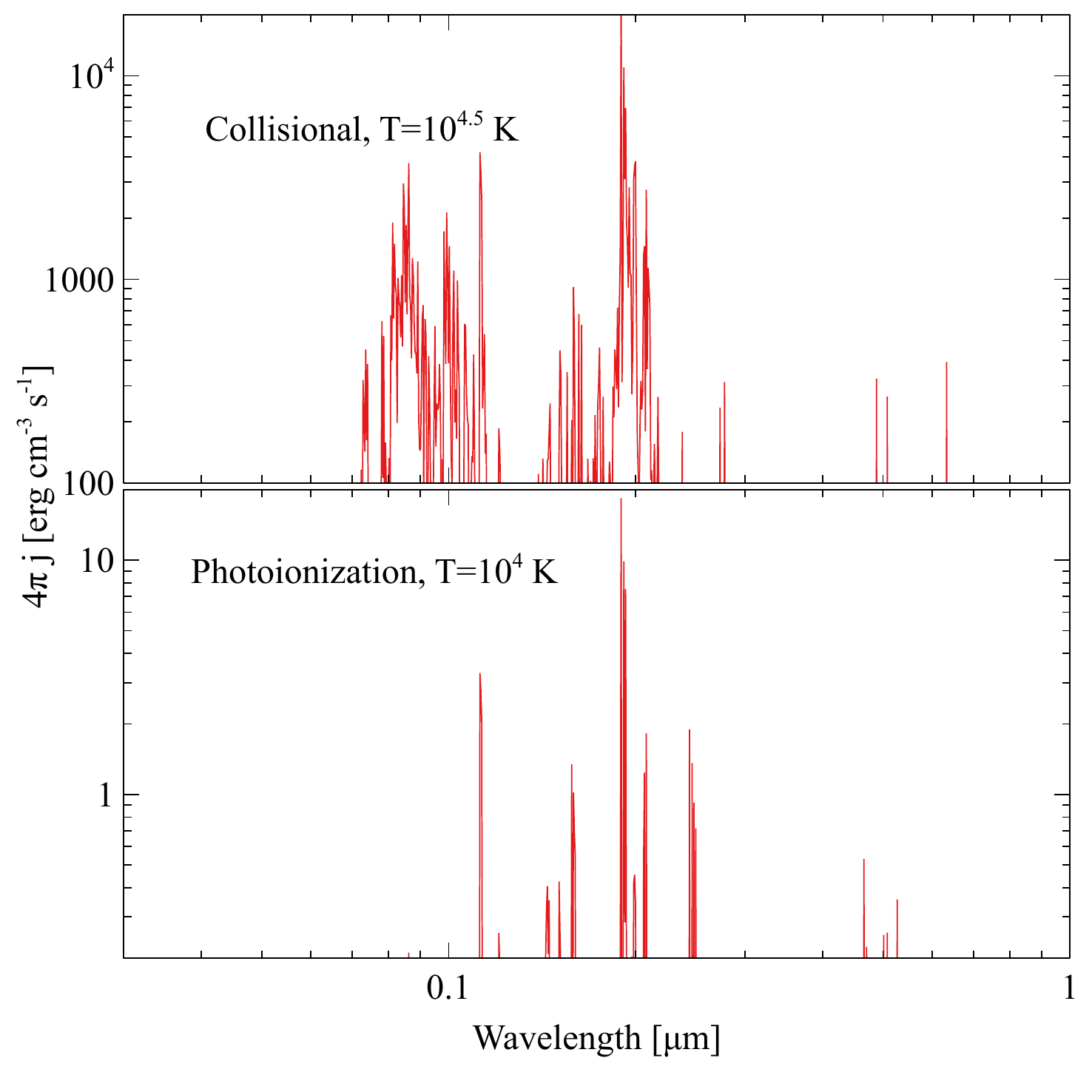}
\end{center}
\caption{The \feiii\ emission spectrum for typical 
photoionization (bottom) and collisional ionization (top) temperatures. 
The gas has only Fe$^{2+}$ and an electron density of $10^{11}\pcc$.
The spectrum is from a unit cell in which the gas experiences only collisional excitation.
The dynamic range of the y-axis is adjusted to
only show lines brighter than one per cent of UV34.
}
\label{fig:Fe3Spectra}
\end{figure}

Figure~\ref{fig:Fe3Spectra} compares the resulting emission spectra of a pure Fe$^{2+}$ gas
with a density of $n_e = 10^{11}\pcc$.
Both spectra are from a `unit cell', a cubic-centimetre of gas, to ensure that the spectrum
is not affected by radiative transfer effects.  
An incident radiation field is not included in the collisional case
so as to ensure a pure collisional model.

The differences in the gas kinetic temperature for the two cases result in spectra of very different form. 
While the UV34 multiplet is strongest in both cases, electron collisions in the much warmer gas of the collisional-ionization model excite higher-energy levels, producing emission at shorter ultraviolet wavelengths. 
 In particular, there is significant emission just below 0.1\,\micron.
 The photoionization case is much cooler so the gas is
 $\simeq$3 dex less emissive (the independent axis in each panel 
 can be compared directly) and short wavelength transitions
 are even weaker.

The difference in predicted \feiii\ emission strength below 1000\,\AA\ is such that existing ultraviolet spectra of luminous quasars can discriminate between a collisional and photoionization origin for the iron emission.
The composite quasar spectrum of \citet[their fig. 5]{2014ApJ...794...75S}, constructed using {\em Hubble Space Telescope}-Cosmic Origins Spectrograph spectra of 159 AGN at $z<1.5$, possesses both high signal-to-noise ratio (S/N) and high resolution ($\simeq$7500). The rest-frame spectrum from the relatively low redshift AGN contributing to the composite is not significantly affected by Lyman absorption from the inter-galactic medium.

\citet{2014ApJ...794...75S} identify \feiii\ emission at 1125\,\AA, 
as predicted by both our photoionized and collisionally ionized models. The quality of the composite spectrum is such that weak emission features with equivalent widths of only a few \AA ngstr\"{o}ms are detectable. 
Examination of their composite, however, confirms there is no 
evidence for \feiii\ emission in the 800-900\,\AA\ region, 
where we would expect to see strong emission if the \feiii\ emitting 
gas was collisionally excited (cf.\,our Fig.~\ref{fig:Fe3Spectra}).

The first conclusion of the paper is, therefore, that there is no evidence for detectable collisionally-ionized \feiii\ emission in the ultraviolet spectra of luminous quasars and a photoionized origin for the \feiii\ emission is strongly favoured. This is consistent with the reverberation of \feiii\ in response to continuum variation reported by  \citet{2018ApJ...862..104M, 2019ApJ...880...96M}.

While collisional-ionization can be ruled out as the physical process responsible for the observed \feiii\ emission, there is an apparent tension between the low emissivity of the \Cloudy \ photoionization model and the strong \feiii\ UV34 emission seen in the \citet{2001ApJS..134....1V} iron emission template. More quantitative measures of \feiii\ emission in the population of luminous quasars can constrain the \Cloudy\ model predictions. 
In the next section, following the definition of the quasar sample to be used,  we therefore make an initial measurement of the strength of \feiii\ UV34 and $\lambda$2075 emission to guide the theoretical investigation presented in Section~\ref{sec:models}.

\section{Observational quasar data}
\label{sec:observations}
\subsection{Quasar sample}
\label{sec:sample}

To characterize the \feiii\ emission in the rest-frame near-ultraviolet spectra of the luminous quasar population we use the quasar catalogue from the fourteenth data release of the SDSS \citep[DR14Q;][]{DR14Q}. Selecting objects with redshifts in the range $1.20 < z < 2.30$ provides rest-frame wavelength coverage over the interval 1700-3000\,\AA\ which includes the \ciii] $\lambda$1909 blend, \mgii\ $\lambda$2800 emission and the complex of \feiii\ lines at $\sim$2075\,\AA.
The reliability of emission line measurements is poor for spectra of the fainter quasars in the DR14Q catalogue and a minimum S/N of 10 per 69\kmps pixel over the wavelength interval 1700-3000\,\AA\ is imposed. 

The resulting sample consists of 26\,501 quasars with redshifts $1.20 < z < 2.30$ and luminosities $10^{45.8} < L_\text{bol}< 10^{48.1}\ergps$. 
The median luminosity of the sample is $L_\text{bol}=10^{46.4}\ergps$.

For comparison with later results, we note that the median luminosity in the sample corresponds to \civ\ emission line lag of $73^{+21}_{-16}$ light-days behind the continuum, assuming the bolometric correction $L_\text{bol} = 3.81 \times L_{1350}$ from \citet{Shen11} and the luminosity-lag relation from \citet{Grier19},
\begin{equation}
    \text{log}_{10}\bigg(\frac{R_\text{CIV}}{\text{light-days}}\bigg) = 0.92 + 0.52 \times \text{log}_{10}\bigg(\frac{L_{1350}}{10^{44}\ergps}\bigg) \pm 0.11
\end{equation}
corresponding to a source-cloud separation of 1.89\e{17}\cm\ for the \civ\ emitting gas.

\subsection{Defining the equivalent width of \feiii\ emission at 2075\,\AA}
\label{sec:2075EW}

A complex of \feiii\ transitions blends together to form a feature at $\sim$2075\,\AA. This feature is relatively isolated, in that it is not blended with emission from other species, and can thus be used to estimate a measure of the strength of \feiii\ emission in a quasar spectrum.

The equivalent width of this emission feature
can be obtained by defining a power-law continuum ($F(\lambda)\propto \lambda^\alpha$) and integrating the emission line flux within a specified wavelength range. The median fluxes in two 10\,\AA-wide intervals, centred on 1975 and 2150\,\AA, are used to calculate the slope, $\alpha$, of the continuum. The continuum-subtracted emission is summed over the wavelength interval 2040-2120\,\AA. Different continuum regions and line-boundaries can be used but the results are not sensitive to the exact wavelengths adopted.

\subsection{Directly detectable UV34 emission}

\begin{figure}
\begin{center}
\includegraphics[clip=on,width=\columnwidth,
keepaspectratio]{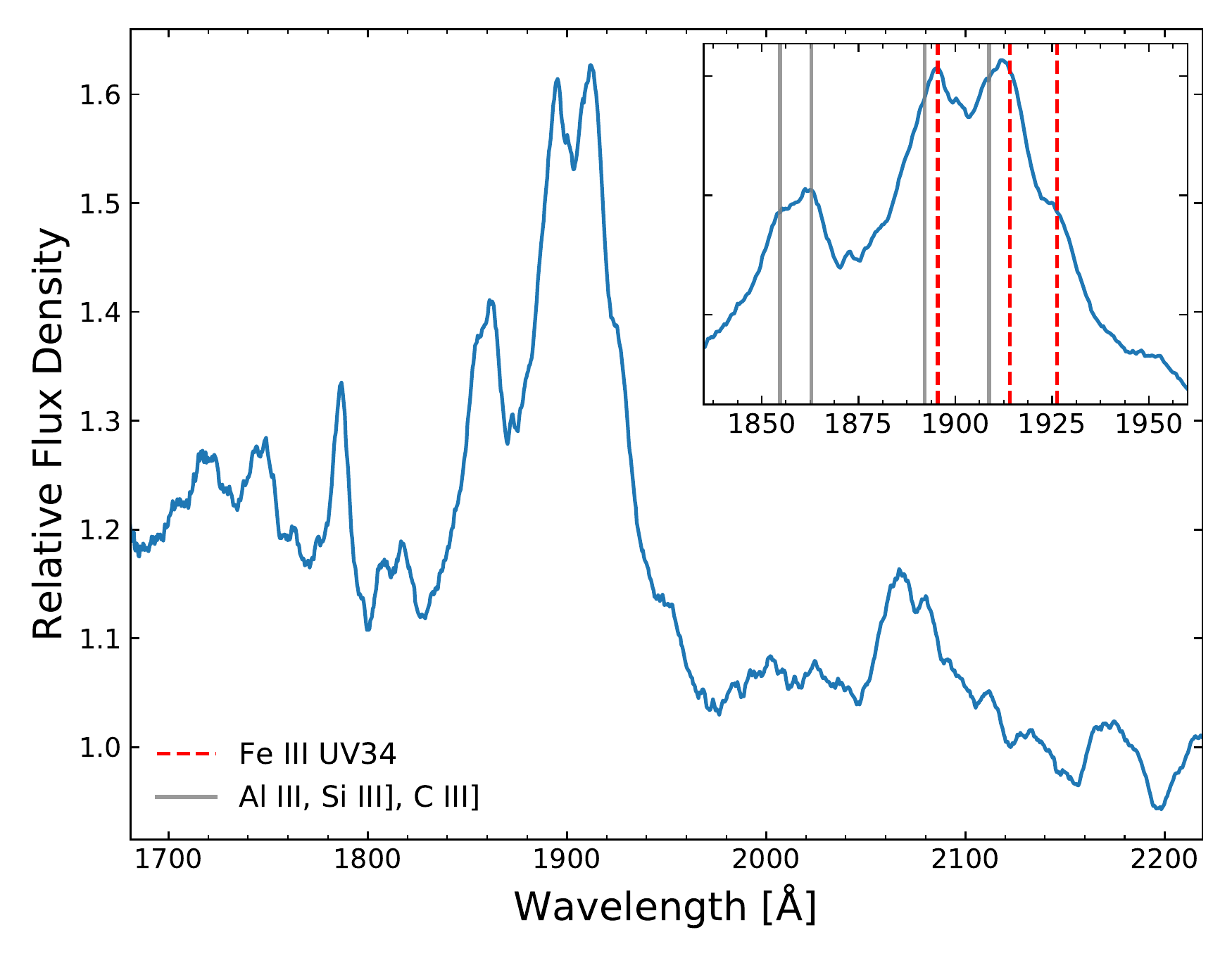}
\end{center}
\caption{Composite of 44 quasars with narrow {\mgii} and and weak \ciii] emission. UV34 emission is  directly detectable, 
and the composite also shows strong \feiii\ emission at 2075\,\AA. The spectrum has been smoothed with a 7-pixel (480\,\kms) window to reduce the pixel-to-pixel noise. 
Inset: the 1909\,\AA\ complex, with the wavelengths of the \ion{Al}{iii}~$\lambda\lambda$1855,1863 doublet, \ciii] $\lambda1909$ and \siiii] $\lambda1892$ lines, and \feiii\ UV34 multiplet marked.}
\label{fig:NarrowComposite}
\end{figure}

The diversity of the morphology of the `1909\,{\AA} complex' among luminous quasars is well-established \citep[e.g.][fig. 11]{Richards11}.
This strong feature consists of the
{\ion{Al}{iii}}\,$\lambda\lambda$1855,1863, \siiii] $\lambda$1892, \ciii] $\lambda$1909 and {\feiii} UV34
emission lines. In most objects, these lines are blended together and it is hard to disentangle the relative contribution of UV34 from the semi-forbidden lines without additional constraints on the line ratios.
However, in objects with narrow emission velocity widths, such as the quasar presented by \citet{Graham96}, 
the individual lines contributing to the 1909\,{\AA} complex deblend and can be directly detected.

Within our sample of quasars it is possible to identify small numbers of 
objects where the full width at half maximum (FWHM) of {\mgii} is narrow
($<$\,2800\,\kms)
and the \ciii] $\lambda$1909 emission is relatively weak.
Figure~\ref{fig:NarrowComposite}
shows a composite of 44 such quasar spectra.
While the definition of their sample differs somewhat,
the quasars contributing to our composite are very similar to
the extreme `Population A' objects discussed by
\citet{2014MNRAS.442.1211M}.
Within this composite, we see that the emission doublet from \ion{Al}{iii}
is detectable at the expected wavelengths, and that 
 \siiii] $\lambda$1892 and \ciii] $\lambda$1909 are very weak.
The \feiii\ UV34 multiplet and 
\ion{Al}{iii}\,$\lambda\lambda$1855,1863 are responsible for the majority of the emission,
demonstrating that in at least some quasars  {\feiii} UV34  is present with high, $\gtrsim5$\,{\AA}, equivalent width.

However, for the majority of quasars,  the  \ion{Al}{iii}, \feiii\ UV34, \siiii] and \ciii] lines are blended together.
At the typical S/N of available large samples of quasar spectra, such as those from SDSS DR14Q,
significant degeneracies exist between measures of the strengths of emission from these species. 
We therefore require additional constraints on the {\feiii} emission lines in order to be able to quantify the strength of UV34 across the whole quasar population.

We note in passing that the composite also shows strong {\feii} UV191\,$\lambda$1787 emission, which is indicative of strong continuum fluorescence (i.e.\,photon pumping, \citealt{2000ApJ...542..644B}),
suggesting that this excitation mechanism may also play an important role in the production of {\feiii} emission.

Before returning to a more extensive investigation of the observational properties of \feiii\ emission in Section~\ref{sec:line_fits}, we first make use of the improved atomic data now implemented in \Cloudy\ to produce theoretical predictions for the strengths of these lines emitted from gas under different physical conditions.

\section{Photoionization calculations}
\label{sec:models}

This section presents predictions of the equivalent widths of a number of ultraviolet lines over a broad range of cloud densities and distances from the black hole.  Our calculations extend to higher densities than conventional BLR grids \citep[e.g.][]{1995ApJ...455L.119B, 1997ApJS..108..401K}. 
We find that the {\feiii} lines trace especially high density gas that is located close to the black hole.  The {\feiii}  energy level diagram 
(Fig.~\ref{fig:Fe3energies})
is highly unusual in that the strongest permitted line, UV34, is a subordinate line that has a different spin than the ground term.  Most ultraviolet lines are instead resonance lines and become strongly thermalised at high densities either because of large optical depths or low critical densities.  They are weak at high densities as a result.  Although the dense clouds do emit in other lines, their contribution is modest compared to emission from lower density components of the BLR. 
As {\feiii} emits only very weakly from lower density gas, we find that {\feiii} is the best tracer of the high density part of the BLR.

\begin{figure*}
\begin{center}
\begin{minipage}{\columnwidth}
\begin{center}
\includegraphics[clip=on,width=\columnwidth,height=0.8
\textheight,keepaspectratio]{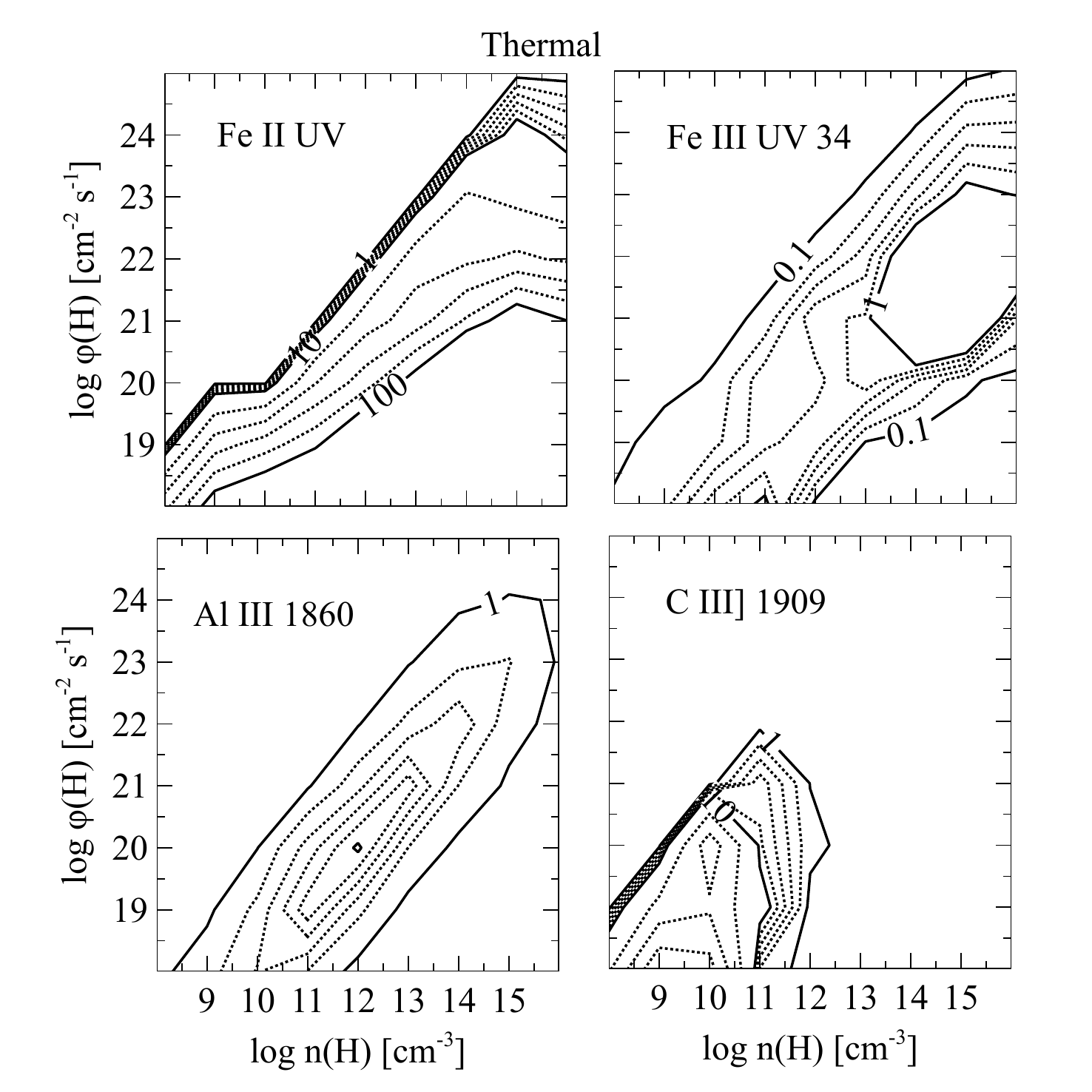}
\includegraphics[clip=on,width=\columnwidth,height=0.8
\textheight,keepaspectratio]{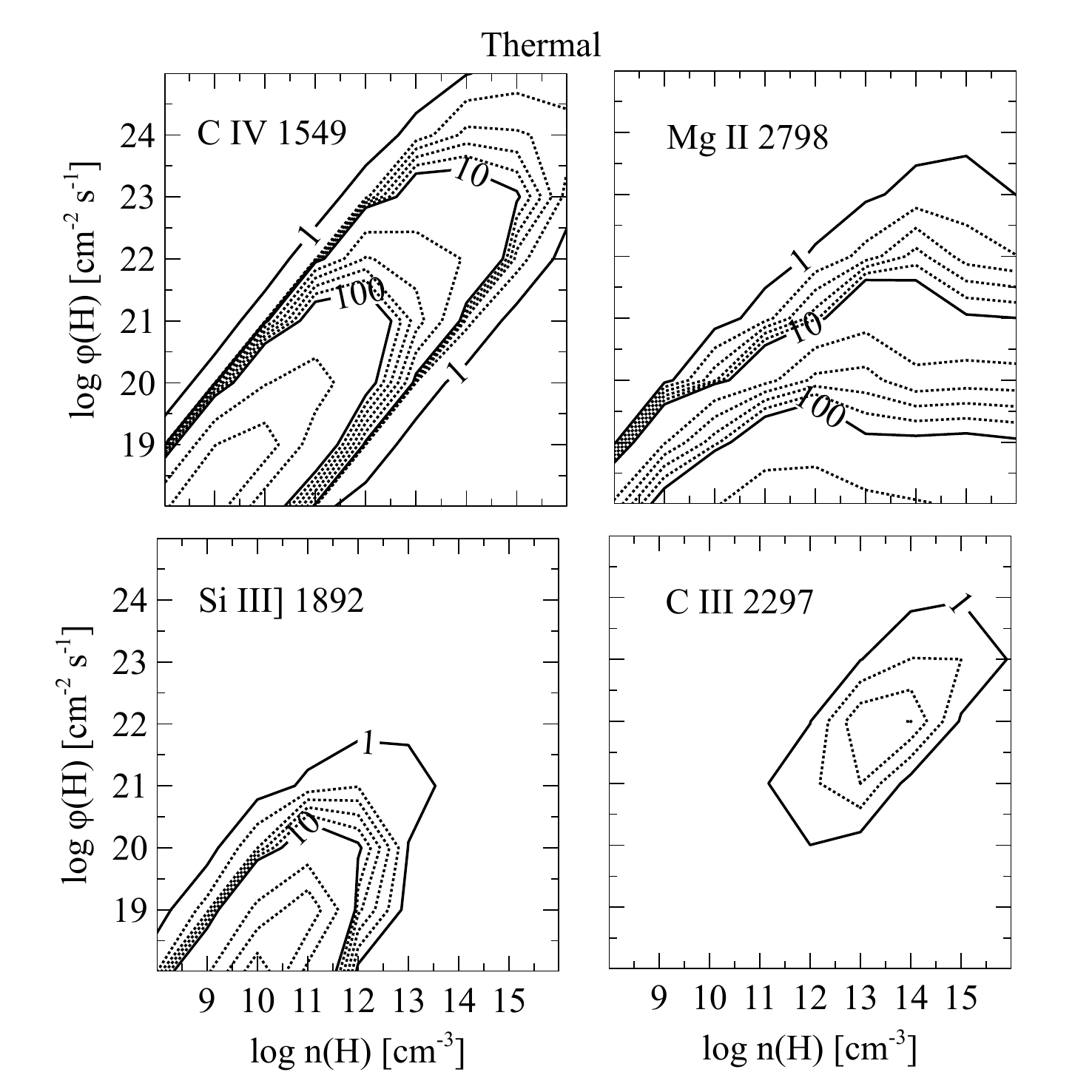}
\end{center}
\caption{The ensemble of possible photoionization models for thermal line widths.
Contours show the predicted equivalent widths in \AA ngstr\"{o}ms relative to the continuum at 1215\,\AA\ for each emission feature,  
assuming a covering factor of unity. The integrated {\feii} 
ultraviolet emission over the interval \mbox{2200-2660\,\AA}
is reported as a single feature.}
\label{fig:thermal1}
\end{minipage}
\hfill
\begin{minipage}{\columnwidth}
\begin{center}
\includegraphics[clip=on,width=\columnwidth,height=0.8
\textheight,keepaspectratio]{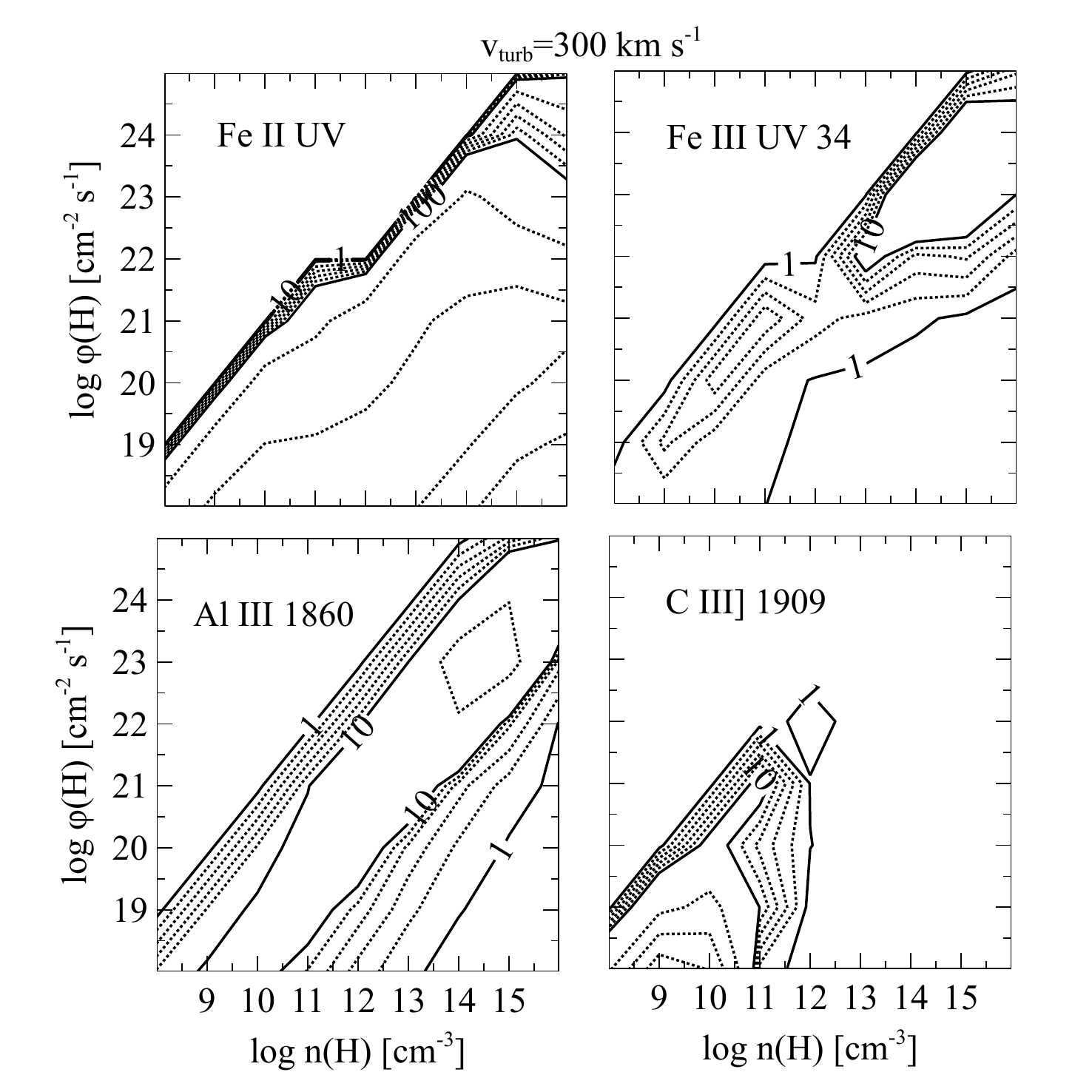}
\includegraphics[clip=on,width=\columnwidth,height=0.8
\textheight,keepaspectratio]{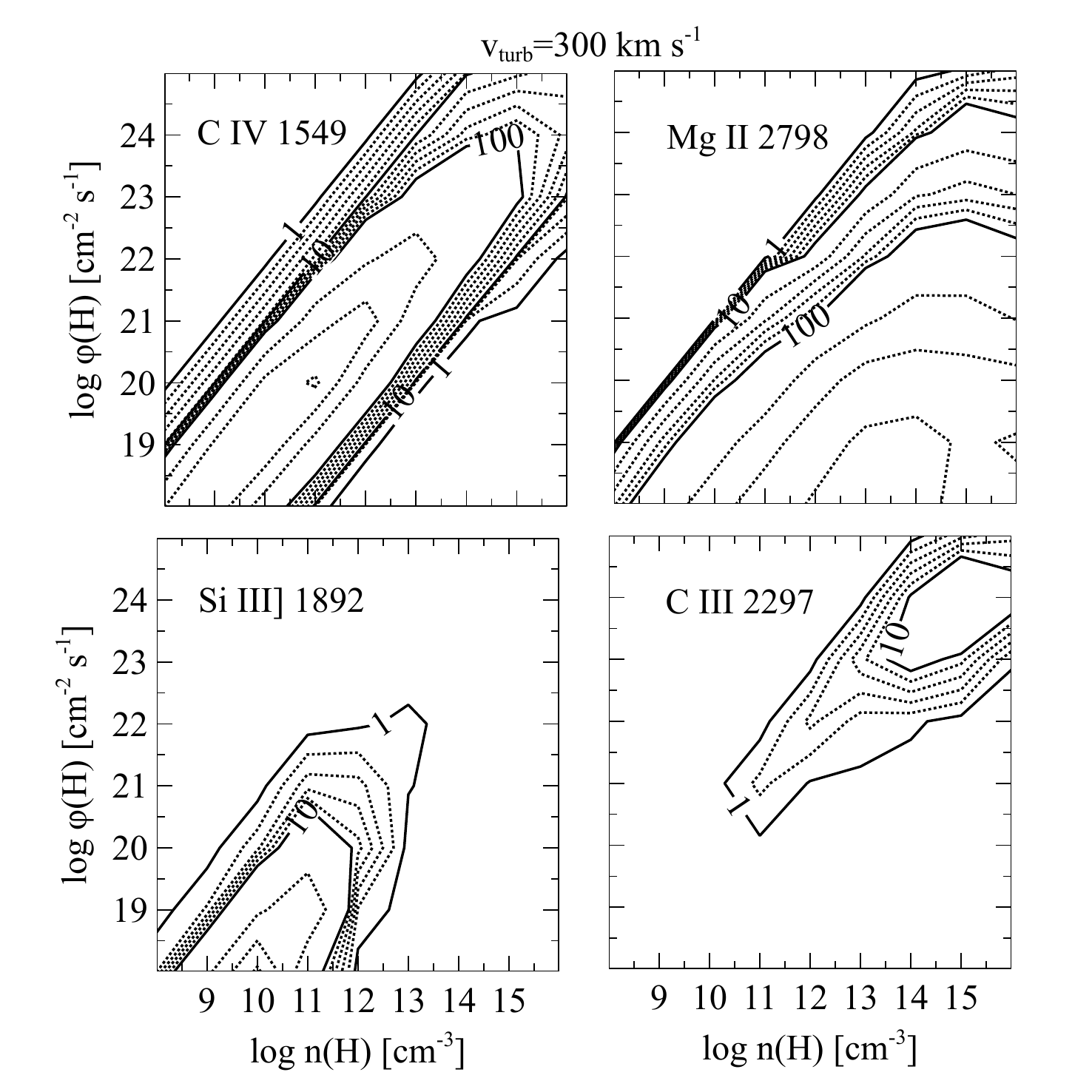}
\end{center}
\caption{The ensemble of possible photoionization models, with microturbulence of 300~km\,s$^{-1}$.
Contours show the predicted equivalent widths in \AA ngstr\"{o}ms relative to the continuum at 1215\,\AA\ for each emission feature,
assuming a covering factor of unity.
The axes are identical to Fig.~\ref{fig:thermal1} and so the contours can be compared directly.
}
\label{fig:turbulent1}
\end{minipage}
\end{center}
\end{figure*}

\subsection{Photoionisation model parameters}

We present a first grid of photoionization models assuming
that lines within the clouds are broadened only by thermal motions, 
the usual assumption
in BLR modeling, and a second grid including a microturbulence of
300\kmps which \citet{2004ApJ...615..610B} found improved the fit
for the \feii\ spectrum.
The effects of including turbulence on other quasar emission lines are described 
by \citet{2000ApJ...542..644B}.

We use the intermediate $L/L_{\rm Edd}$ SED described by 
\citet{2012MNRAS.425..907J}.
We assume solar abundances and a
cloud column density of $10^{23} \pscm$, typical assumptions for BLR clouds.
With these assumptions the remaining parameters are the flux of ionizing
photons striking the cloud, $\phi({\rm H}) [\pscm \ps]$, and the hydrogen density
$n_H [\pcc]$.  
In Figs. \ref{fig:thermal1} and \ref{fig:turbulent1} we vary these parameters over
a broad range.
Plots similar to these are presented in 
\citet{1995ApJ...455L.119B},
\citet{1997ApJS..108..401K}, and
\citet{2004ApJ...615..610B}.

The following features are presented:
 \ion{C}{Iv} $\lambda$1549, 
 \ion{Al}{III} $\lambda$1860,
 \ion{Si}{III}] $\lambda$1892,
 \ion{C}{III}] $\lambda$1909,
\ion{C}{III} $\lambda$2297,
 \ion{Mg}{II} $\lambda$2798,
{\feii} ultraviolet emission over the interval 2200-2660\,{\AA} (`{\feii} UV'),
and \feiii\ UV34.
All features are shown as equivalent widths in \AA  ngstr\"{o}ms expressed relative to the
continuum at 1215\,\AA,
and assuming $4\pi\sr$ coverage of the ionizing source, i.e. a cloud covering factor of unity. 
In the case of multiplets or blends we predict the total equivalent width.  
However, the line luminosity or equivalent width depends on the cloud
covering factor \citep{2006agna.book.....O}, and so
these predictions need to be multiplied by that factor,
which is typically taken to be of order 20 per cent.
Adopting this value, an observed \feiii\ UV34 equivalent width of 5\,\AA\
requires a predicted equivalent width of order 25\,\AA.

The two axes for these contour plots have simple physical meanings.
The flux of ionizing photons $\phi({\rm H})$ is related to the total ionizing
photon luminosity, $Q({\rm H})$ [\ps] and the distance of the cloud from
the continuum source $r$ by
\begin{equation}
    \phi(\rm{H}) = \frac{Q({\rm H})}{4\pi r^2} [\pscm \ps]
\end{equation}
so the vertical axis is related to the cloud separation
from the centre,  $\phi \propto r^{-2}$ with lower regions
representing larger radii.  
The typical bolometric luminosity of our sample is $10^{46.4} \ergps$. 
This, assuming the middle SED of \citet{2012MNRAS.425..907J}, corresponds to a
photon luminosity of $Q({\rm H}) = 1.96\e{56}\ps$.
For reference, \civ\ emitting clouds, adopting a lag of 73 light days,
have 
\begin{equation}
\phi({\rm H}) = 4.36\e{20} r_{73.6}^{-2} [\pscm \ps],
\end{equation}
where $r_{73.6}$ is the radius in light days 
and we assume that the radiation field falls off as an inverse square law. 
Lower-ionization photoionized clouds that emit strongly in the
ultraviolet and optical have kinetic temperatures in the neighbourhood of
1-2\e{4}\K\ so the horizontal axis is an approximate surrogate
for the gas pressure.
We see that the source-cloud separation varies by nearly 4 dex while
the gas pressure varies by 8 dex.

Intermediate ionization lines such as \ion{C}{iv} and \ion{Al}{III} peak along a 
diagonal corresponding to particular values of the ionization parameter
\begin{equation}
U=\frac{\phi({\rm H})}{c\, n_H}.
\end{equation}
Very low ionization lines such as \feii\ and \ion{Mg}{II} do not show such a peak
but rather favour lower $\phi({\rm H})$ and larger $r$.

\subsection{Implications of the \feiii \ equivalent width; a turbulent dense medium}

The energy level diagram shown in Figure \ref{fig:Fe3energies} suggests that
UV34 is an odd choice for the strongest \feiii\ line.  
The multiplet has a high excitation potential, with its
upper level at 
82\,000 wavenumbers (10.2\,eV).
Direct excitation from the ground term is via forbidden transitions
and so is not very efficient.  
Tests show that efficient excitation of UV34 is via a two-step process, with
excited quintet levels playing an intermediary role.  

Several {\feiii} levels are close enough to 
the Ly\,${\rm \alpha}$ energy of 82 259 wavenumbers
for Ly\,${\rm \alpha}$ pumping to be significant.
This is included in {\Cloudy} as a general line excitation
process and completes with direct continuum
photoexcitation.  
Both processes will photoexcite the septets
since these are connected to ground by
an allowed electric dipole transition.
However, the fact that the spectrum is sensitive to the
turbulence, which affects continuum 
pumping more than line overlap,
shows that continuum fluorescence is more important than 
Ly\,${\rm \alpha}$ pumping.

\begin{figure}
\begin{center}
\includegraphics[clip=on,width=\columnwidth,height=0.8
\textheight,keepaspectratio]{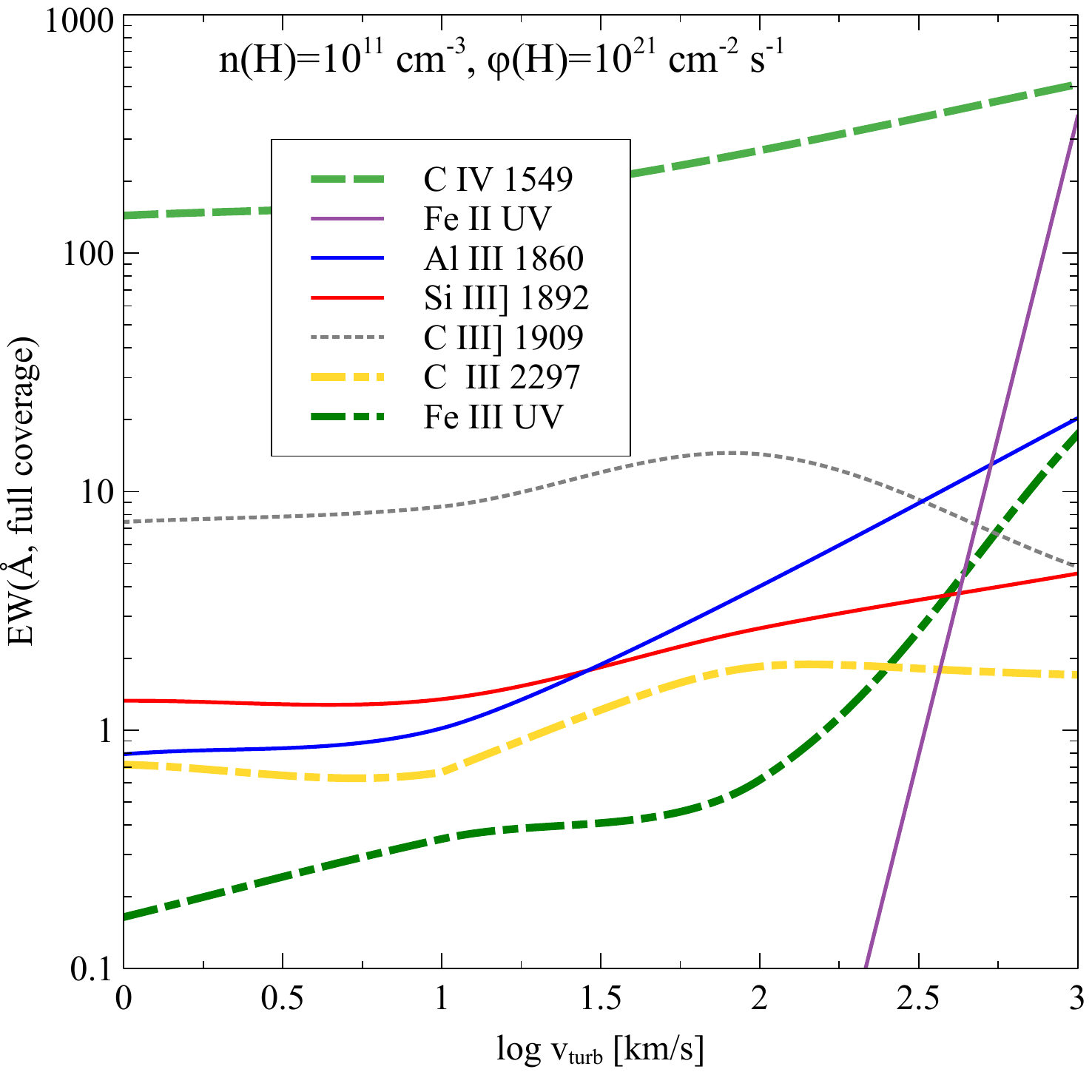}
\end{center}
\caption{The effects of varying turbulence upon the equivalent widths of several
prominent lines.
Both \feiii\ and \feii\ are predicted to produce much stronger emission as the turbulent velocity parameter increases.
}
\label{fig:VaryTurb}
\end{figure}

The effects of varying the turbulence are shown in Fig. \ref{fig:VaryTurb}.
Increasing the turbulence increases the importance of continuum pumping
because the line width increases and so do the number of continuum photons
that can be absorbed by a transition \citep[Section 2.1 of][]{1992ApJ...389L..63F}.
Physically, increased continuum pumping increases the population of excited quintets
which can then undergo a collisional transition to the septets.

We note that, as discussed in \citet{2004ApJ...615..610B}, the turbulence implemented in \Cloudy\ describes any situation in which the wavelengths of the emitting lines are shifted or broadened  over a physical distance corresponding to the mean free path of a continuum photon. Such velocity differences could naturally arise from differential rotation or instabilities within the accretion disc. 
We see no evidence for kinematic structure within the \feiii\ lines to suggest that the \feiii\ emitting material is itself outflowing, and so we do not believe that the turbulence needed to account for the observed strength of the \feiii\ emission is due to velocity gradients in ordered outflows as would be expected if the \feiii\ emitting material was entrained in a wind off the accretion disc.

\begin{figure}
\begin{center}
\includegraphics[clip=on,width=\columnwidth,height=0.8
\textheight,keepaspectratio]{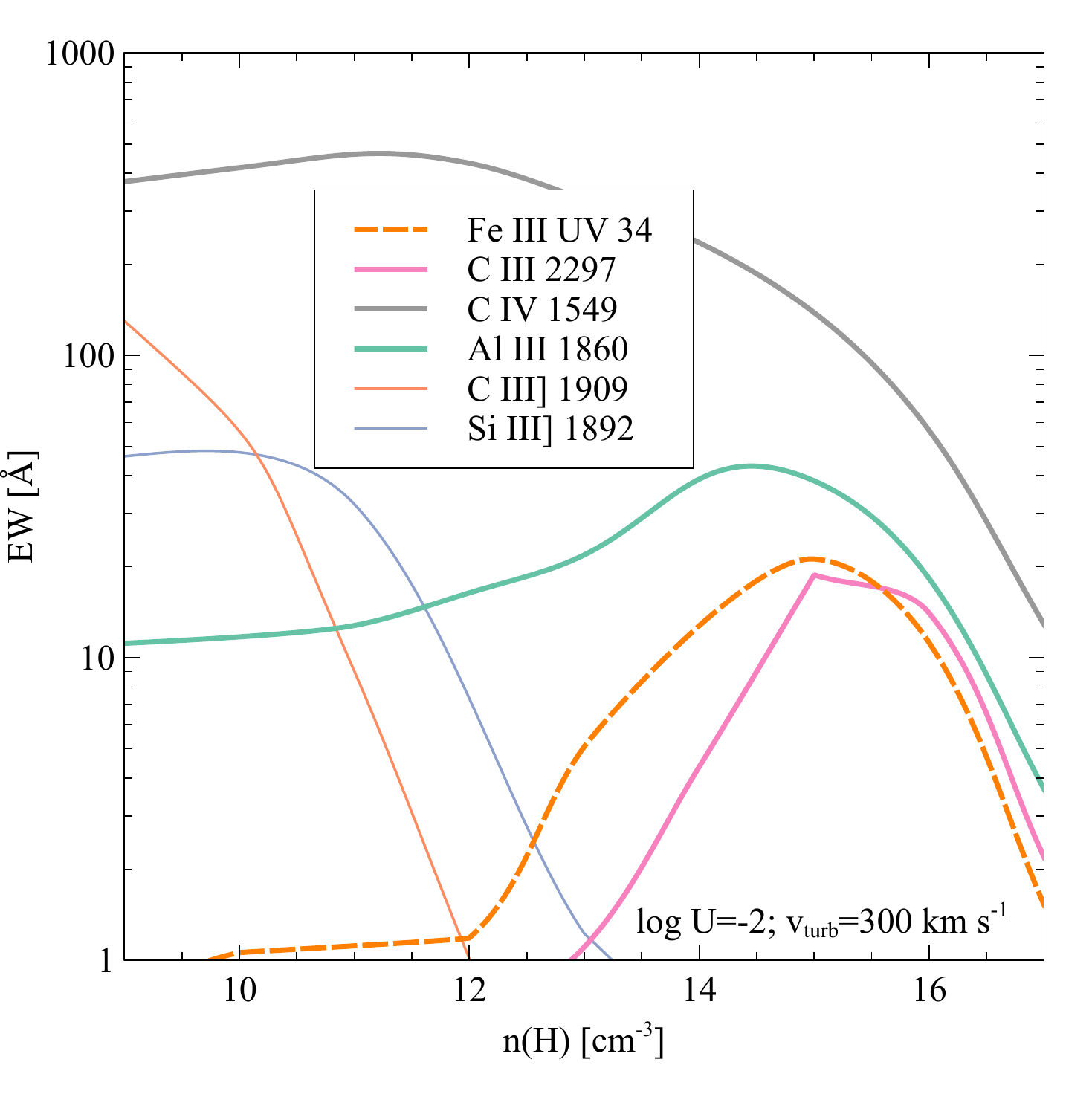}
\end{center}
\caption{The effects of varying hydrogen density along the ridge of peak \civ\ emission strength,
keeping the ionization parameter constant.
\ciii]~$\lambda1909$ and \siiii]~$\lambda1892$ produce very little emission above $10^{13}\pcc$, where both \feiii\ UV34 and \ion{Al}{iii} $\lambda 1860$ are predicted to peak. \ciii~$\lambda 2297$ is predicted to be strong above $10^{14}\pcc$, but this line is not observed in quasar spectra.
}
\label{fig:varyn}
\end{figure}

Adopting a value of $v_{\rm turb}=300\kmps$ for the microturbulence parameter, we find that the predicted equivalent width of \feiii\ emission is still significantly weaker than that observed in quasar spectra. 
Varying the metallicity of the emitting gas has an insufficient effect on the predicted equivalent width for the strength of \feiii\ in AGN to be explained by non-solar iron abundances.
However, as shown in Figs.~\ref{fig:thermal1} and \ref{fig:turbulent1}, the \feiii\ emission peaks at a density of $\simeq$$10^{15}\pcc$, irrespective of the microturbulence.
In Fig.~\ref{fig:varyn}, we show how the equivalent widths of various lines change as a function of density, taking a diagonal slice through the $\phi({\rm H})-n_H$ plane corresponding to the ridge of peak \civ\ and \ion{Al}{iii} emission and constant ionization parameter.  

The only way in which we can account for the observed strength of the \feiii\ emission is through turbulent, high density gas with $n_H>10^{13} \pcc$. Taken with the upper limits on the density of the gas which is producing the semi-forbidden lines such as \ciii]\,$\lambda$1909 in quasar spectra, this is strong evidence for  multiple components of differing densities contributing to the region which is  producing the broad emission lines in AGN.

\subsection{How dense is the turbulent dense medium?}
\label{sec:ciii}

\begin{figure}
\begin{center}
\includegraphics[clip=on,width=\columnwidth,height=0.8
\textheight,keepaspectratio]{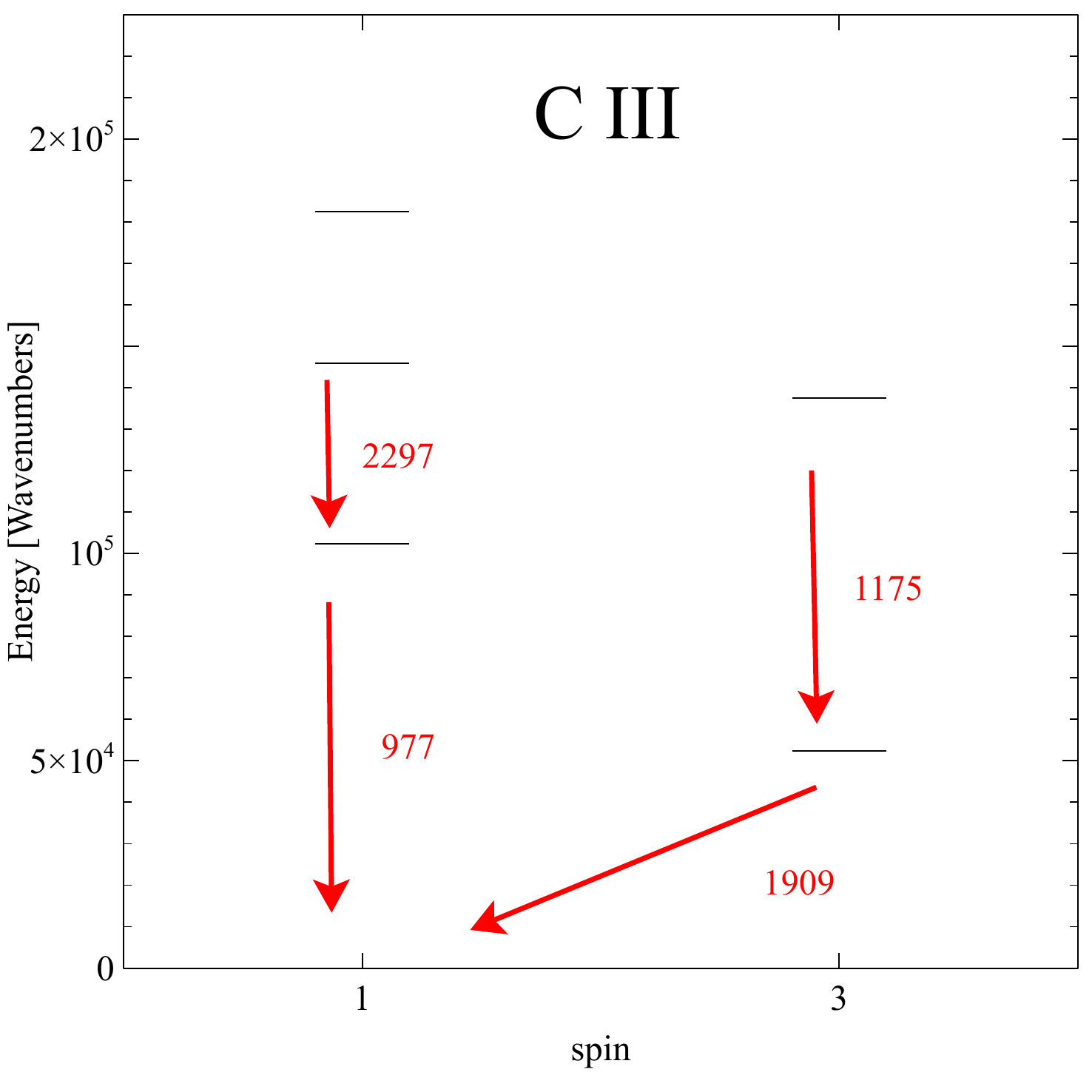}
\end{center}
\caption{The \ciii\ model implemented in Cloudy. 
Only the lower levels are shown to keep the energy scale the same as Fig.~\ref{fig:Fe3energies}.
The strong \ciii\ lines are indicated.}
\label{fig:C3energies}
\end{figure}

In the previous subsection, we have shown that emission from a turbulent medium with density $n_H>10^{13} \pcc$ is required to reproduce the observed \feiii\ line strengths. To better constrain the density of this medium, we use the predicted strengths of the \ciii\ emission lines shown in Fig.~\ref{fig:C3energies}.

As shown in Fig.~\ref{fig:varyn}, the strength of the semi-forbidden \ciii]\,$\lambda$1909 line drops drastically above $n_H\simeq10^{12} \pcc$. We show in Section \ref{sec:line_fits} that the \ciii]\,$\lambda$1909 emission in our sample of SDSS quasars is independent of the \feiii\ UV34 emission, consistent with emission from at least two distinct locations around the black hole.

The \ciii\,$\lambda$2297 line is predicted to be strong when the density is above $n_H\simeq10^{14} \pcc$  (Fig.~\ref{fig:varyn}). 
However, to the best of our knowledge, the line is not detected in the spectrum of any quasar. In particular, \ciii\,$\lambda$2297 is not present in the high signal-to-noise composites of \cite{VandenBerk01} and \cite{Francis91}, or the ultraviolet spectrum of I\,Zw\,1 presented in \cite{2001ApJS..134....1V}.
We therefore use the observed lack of \ciii\,$\lambda$2297 emission to constrain the density of the \feiii\ emitting gas to lie in the range $n_H=10^{13-14} \pcc$.

\section{Comparison with observations}
\label{sec:line_fits}

With theoretical predictions from \Cloudy\ in hand, we can now  use the full sample of spectra described in Section \ref{sec:sample} to test our preferred model for the dense turbulent medium required to reproduce the observed \feiii.

\subsection{Measuring \feiii\ UV34 in quasar spectra}

\begin{figure}
\begin{center}
\includegraphics[clip=on,width=\columnwidth,
keepaspectratio]{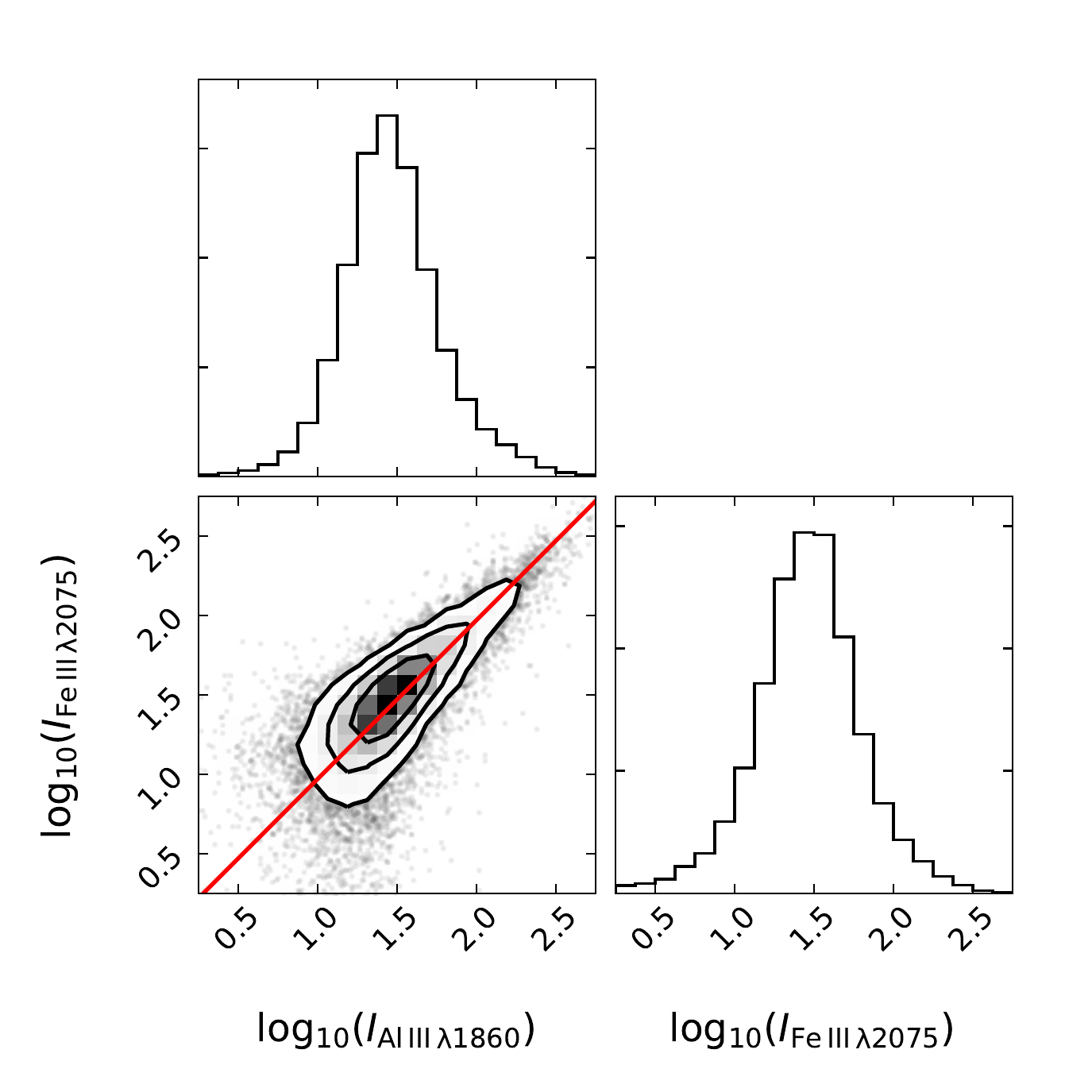}
\end{center}
\caption{The observed intensities of the \ion{Al}{III}\ doublet  and \feiii\ $\lambda$2075 complex in units of $10^{-17}$\ergps\pscm.  In red is the predicted ratio from the dense turbulent \Cloudy\ model, which agrees very well with the slope of the observed correlation. The observed emission is consistent with the line ratio expected if both lines are produced in the same dense turbulent medium.}
\label{fig:Al3}
\end{figure}

As shown in Figs.~\ref{fig:VaryTurb} and \ref{fig:varyn}, the dense turbulent medium will also produce a significant amount of \ion{Al}{iii} doublet emission at $\lambda\lambda$1854.72,1862.79.  For each of the 26\,501 quasars in our sample, we fit the \ion{Al}{iii} doublet with two Gaussians, with the ratio of the lines constrained to be $2.3866:1.9162$, and compare with the strength of the \feiii\,$\lambda$2075 complex described in Section \ref{sec:2075EW}. The results are shown in Fig.~\ref{fig:Al3}: the strength of the \ion{Al}{iii} doublet clearly correlates with the strength of the \feiii\,$\lambda$2075. 
The ratio of the line intensities of the two complexes is in agreement with the predictions of our dense turbulent model.

Encouraged by this, we fit the $\lambda$1909 complex with a model which describes the relative contribution of the aluminium, iron, carbon and silicon lines.

For each object, a power-law in flux is defined at 1800 and 2020\,\AA\ and subtracted from the spectrum. 
The wavelength region 1820-1920\,\AA\ is then fit with a sum of seven Gaussians described by four free parameters: (i) the amplitude of the \ion{Al}{iii} and \feiii\ UV34 component, (ii) the amplitude of the \ion{Si}{iii}] line, (iii) the amplitude of the \ciii] line, and (iv) the velocity width of the lines.
The \ion{Al}{iii} doublet and \feiii\ UV34 triplet are constrained to have the line ratios
\begin{multline}
    \lambda 1854.72:\lambda 1862.79:\lambda  1895.46: \lambda 1914.06: \lambda 1926.30 
    \\ = 2.3866: 1.9162 : 0.9910: 0.7538: 0.6945
\label{eq:ratios}
\end{multline}

\noindent
as predicted for a dense, microturbulent, photoionized gas\footnote{$\nu=300$~\kms, $n_H=10^{14}\pcc$, $\phi=10^{23}\pscm\ps$}
with solar abundances.
In theory the \ion{Si}{iii}] and \ciii] lines would be expected to have narrower profiles than the \ion{Al}{iii} and \feiii, as they are coming from gas which is less dense and thus located further  from the black hole.
In practice the S/N of the spectra are such that allowing the velocity widths of the lines to vary independently  leads to an over-fit to the data. We therefore constrain the velocity widths of all lines in our fitting routine to be equal, which gives a more robust estimate of the equivalent width of each line in the complex. 
Examples of our best fitting models are presented in Appendix~\ref{sec:example_fits}.

\subsection{Implications for C III] and Si III]}

The improved atomic data and \Cloudy\ model presented above now allow a more certain estimate of the UV34 multiplet to be made in any given quasar spectrum, even when the multiplet is completely blended with other lines. This allows, for the first time, an accurate measurement of the \ciii] and \ion{Si}{iii}] emission to be made in objects with a significant contribution from \feiii\ UV34.

\begin{figure}
\begin{center}
\includegraphics[clip=on,width=\columnwidth,
keepaspectratio]{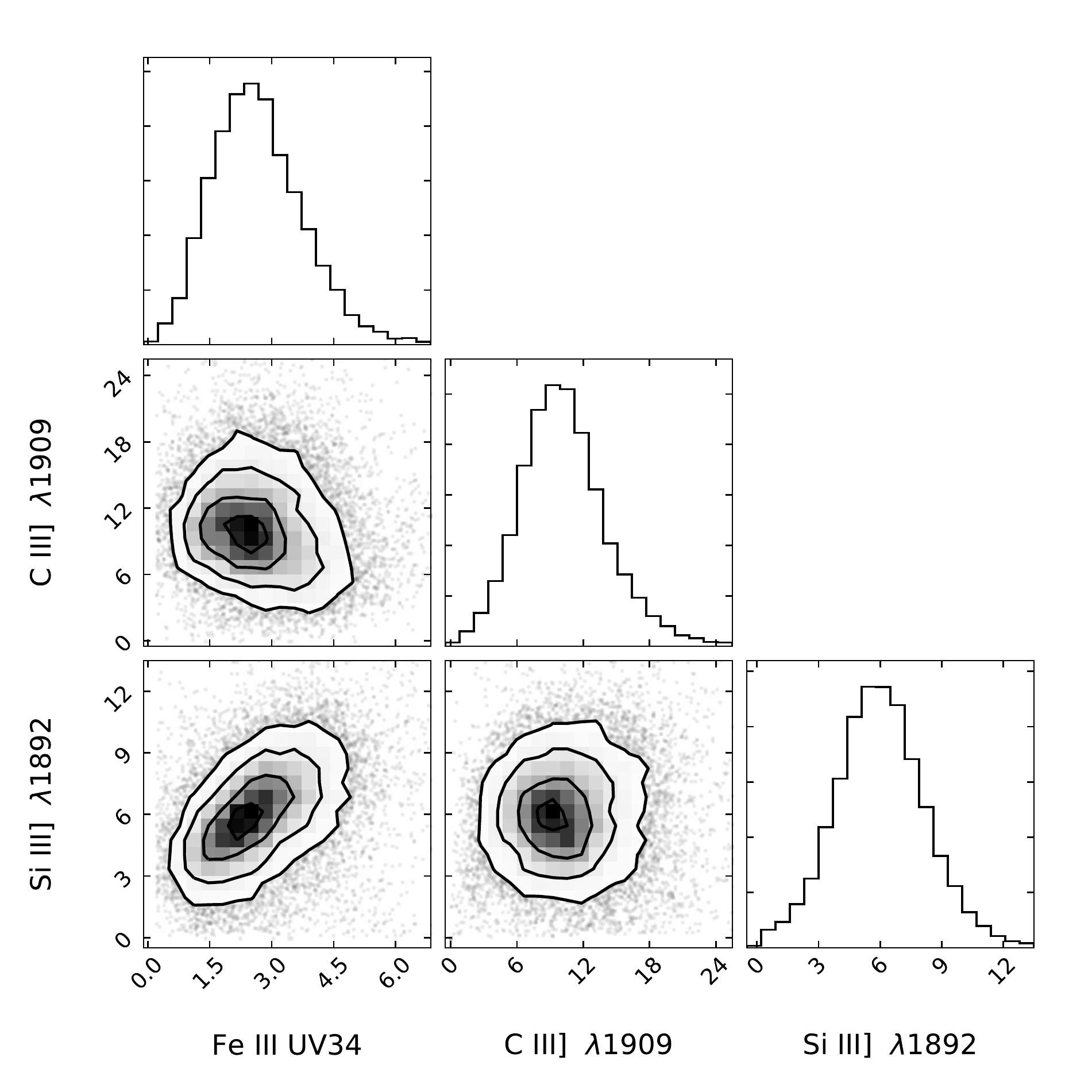}
\end{center}
\caption{Observed equivalent widths in \AA ngstr\"{o}ms for three components of the $\lambda$1909 complex, relative to the continuum at 1900\,\AA. When the \feiii\ UV34 emission is incorporated into the fit to the complex, we find that the \ciii] emission strength is uncorrelated with that of \siiii].
}
\label{fig:fits}
\end{figure}

Using our dense turbulent model for \feiii\ emission to fit the $\lambda$1909 complex, with independent contributions from \ion{Si}{iii}] and \ciii], we measure the equivalent widths of the
\ion{Si}{iii}], \ciii] and \feiii\ lines in our sample of SDSS quasars.
The results are shown in Fig.~\ref{fig:fits}. 
We find that the \feiii\ emission correlates with the \ion{Si}{iii}], but that the \ciii] emission is essentially uncorrelated with either species.
This is consistent with the expected behaviour of \ciii] $\lambda$1909, which must be coming from gas of a significantly lower density than that which is emitting the \feiii\ (see Section \ref{sec:ciii}).

The \ion{Si}{iii}] strength does however correlate with the strength of the \feiii\ UV34, suggesting that some of the observed silicon emission could also be coming from the dense turbulent component.
The observed \ciii]:\siiii] ratio thus correlates with \feiii\, and could therefore be an indicator of the hardness of the SED which is illuminating the broad line region, as suggested by \citet{Casebeer06} and \citet{Richards11}.

The median equivalent widths of emission in our sample are 2.58, 9.95, 5.97, and 4.43 {\AA} for {\feiii} UV34, \ciii], \siiii], and {\ion{Al}{III}} respectively; the average strength of {\feiii} UV34 is $\approx$0.16 times the combined strength of \ciii] and \siiii]. 83 per cent of the quasars in our sample have UV34 fluxes which are greater than 0.1 times the sum of flux in \ciii] and \siiii].
This highlights the need for accurate modeling of UV34 when attempting to measure  \ciii] and \siiii] line properties.

\subsection{Broad absorption line quasars}

The absorption line properties of the 19\,590 quasars in our sample with $z>1.56$, where spectral coverage of the \civ\ emission line is available, are classified as part of the \citet{Rankine20} investigation into broad absorption line (BAL) quasars.
The classification of objects as BAL quasars from this subsample produces 16\,148 non-BAL and 3048 high-ionization BAL quasars. We exclude 394 low-ionization BALs with broad \ion{Al}{iii} troughs from subsequent analysis.

Following the same scheme as described above, we find no significant difference between the 
high-ionization BALs and non-BALs, in that the 
high-ionization BALs also require a high density, turbulent component  to account for the \feiii\ emission in their spectra.

\subsection{Comparison with Vestergaard \& Wilkes (2001)}

In this section we compare our \feiii\ UV34 multiplet with those presented in the templates of  \citet{2001ApJS..134....1V}.

\cite{2001ApJS..134....1V} discuss the blending of UV34 with the \ciii] and \ion{Si}{iii}] lines and the degeneracies arising when attempting to fit multiple overlapping lines simultaneously due to the non-orthogonality of Guassian functions (their section~5.1).
Due to the lack of atomic data at the time, they were unable to place theoretical constraints on the UV34 line ratios and thus provided a range of ratios which were consistent with the observed emission. 
Their published ratios range from the optically thick case where the lines thermalise at 1:1:1, to their preferred values of  0.375:1:0.425
based on the observed strength of the $\lambda$1914 line in I\,Zw\,1.

Using the new atomic data and the dense turbulent \Cloudy\ model presented above, we are now able to predict the ratios of the $\lambda\lambda$1895.46,1914.06,1926.30 \feiii\ lines in quasar spectra.
Assuming a dense, microturbulent, photoionized gas
with solar abundances, we find UV34 line ratios of 1.316:1:0.921.

Our preferred \feiii\ model therefore predicts stronger $\lambda$1895 emission relative to $\lambda$1914 than any of the models presented by \citet{2001ApJS..134....1V},
but we believe our line ratios are still consistent with those they observe in I\,Zw\,1 within the uncertainties they discuss.

\section{Discussion}
\label{sec:discuss}

\subsection{High Densities}

It is perhaps surprising to see that significant line emission from
{\feiii} and {\ion{Al}{iii}} is predicted from gas  illuminated by such high ionizing fluxes.
Our favoured model for the {\feiii} emitting gas has an ionizing photon flux $\phi=10^{23}\pscm\ps$,
which suggests that this part of the BLR must lie very close to the central black hole.
While this is consistent with the existing reverberation-mapping studies in the literature
(see Section{~\ref{sec:intro}}), we would expect this locale to have large Keplerian speeds  
- the 4\,dex difference in $\phi$ gives a 2\,dex difference in radius and hence ${v}_{\rm \,Kepler}$ a factor
10 greater  compared to the gas which is most likely emitting \ciii]. 

We would expect such an extreme difference in the velocity width of the 
emission lines to be easily observable, however, no such difference is 
seen in the objects in our sample. 
The spectral resolution and S/N constraints are such that, while we cannot
rule out variation between the velocity widths of each species within 
individual objects at the level of tens of per cent, 
the \siiii] and \ciii] lines are no more than a factor 2 narrower 
than the {\ion{Al}{iii}} lines in any given object in our sample.

One explanation for this is if the dense turbulent gas which is emitting {\feiii} is primarily undergoing equatorial motion.
For a Type-1 quasar, this  would mean that the majority of the motion of the line emitting gas is in the plane of the sky, and the observed velocity dispersion along the line of sight is  smaller than the Keplerian speed.
This would suggest that the dense media we require to explain the strength of the observed {\feiii} emission are found in quasi-ordered flows in the equatorial regions of the AGN.

\subsection{Correlations with \civ\ properties}

\begin{figure}
\begin{center}
\includegraphics[clip=on,width=\columnwidth,
keepaspectratio]{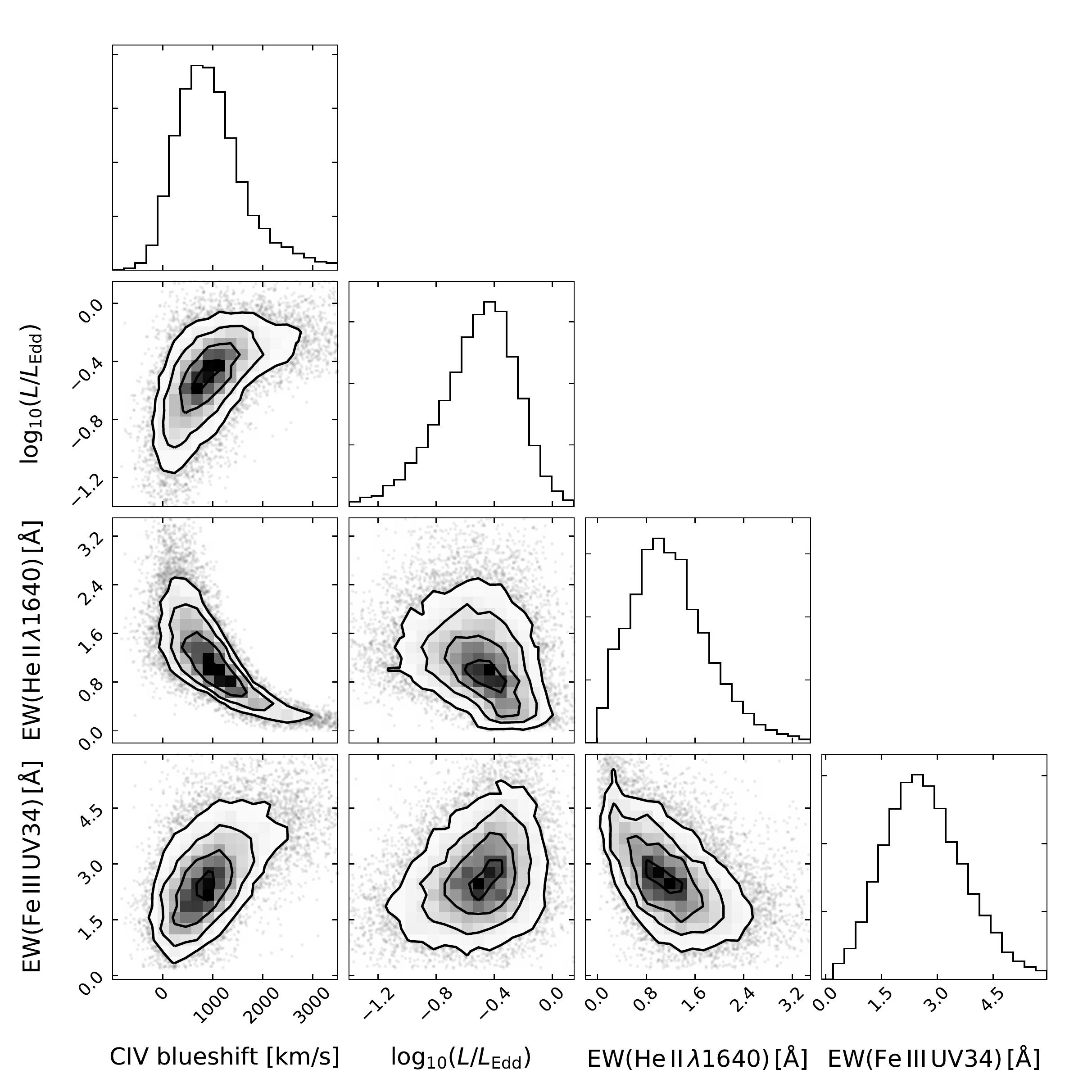}
\end{center}
\caption{The equivalent width of {\feiii} UV34 emission,
shown against the observed blueshift of the {\civ} emission line \citep[as defined in][]{Rankine20},
the Eddington ratio, and the equivalent width of  {\ion{He}{ii}}\,$\lambda$1640 emission. 
The strength of {\feiii} increases with the signature of stronger nuclear outflows, 
which might be driven by higher $L/L_{\rm Edd}$.
However, the strengths of {\feiii} and {\ion{He}{ii}} anticorrelate, suggesting that the link between {\feiii} and {\civ} could also be driven by variation in the SED which is illuminating the BLR.}
\label{fig:C4}
\end{figure}

The high-ionization \civ\ $\lambda 1549$ emission line exhibits a large range of kinematic morphologies,  from strong, `peaky' emission at the systemic redshift which is believed to be dominated by emission from gas in virial equilibrium, to weaker and highly blueshifted emission tracing outflowing material. However, the physical mechanisms which set the balance of these emitting components are not well understood, despite the fact that the \civ\ morphology is known to be closely tied to other parameters such as the Eddington ratio and the hardness of the SED which is ionizing the BLR \citep{Richards11, Rankine20}.
Here we note certain correlations using the results of the modelling of the $\lambda 1909$ complex described in Section~\ref{sec:line_fits}. 

The measured strength of the \feiii\ and \ion{Al}{iii} component tends to increase as the blueshift of the \civ\ line increases (Fig.~\ref{fig:C4}), suggesting that objects with stronger wind emission are likely to display stronger emission from the dense turbulent medium.
This is consistent with the results from the sample of seven objects studied by
\citet{1996ApJ...461..664B}.
However, the \feiii\ line profiles  do not themselves show any  evidence for outflows, and so we do not believe that the dense ($\simeq$$10^{14}\pcc$) \feiii\ emitting gas is being accelerated within a wind. 

One possible explanation for the correlation between the strength of the {\feiii} UV34  emission and the increasing blueshifted {\civ} emission, tracing outflowing lower density material, might be the range of Eddington fractions within the population, which is one of the commonly quoted drivers of variation in  {\civ} line properties.  Quasars with larger Eddington fractions potentially `puff up' the inner regions of the accretion disc \citep[e.g.][]{2019A&A...630A..94G}, leading to a larger volume of high density line emitting gas close in to the black hole. The increased Eddington fraction also leads to stronger radiation line-driven winds. 
However, inferring $L/L_{\rm Edd}$ from single-epoch spectra is fraught with large systematic uncertainties, deriving both from the use of a single bolometric correction to estimate $L_{\rm bol}$ from a monochromatic luminosity, and also the use of a virial factor to estimate the black hole mass from a single emission line velocity width.
Here we use the FWHM of the {\mgii} $\lambda$2800 line and the single-epoch virial estimator described by
\citet{Vestergaard09}
to estimate black hole masses, and the monochromatic luminosity at 3000\,{\AA} to estimate $L_{\rm bol}$ and hence $L/L_{\rm Edd}$. This quantity is also shown in Figure~\ref{fig:C4},
along with the observed equivalent width of the {\ion{He}{ii}} $\lambda$1640 emission line.

As expected, we recover the correlation between $L/L_{\rm Edd}$ and {\civ} blueshift.
We also find that objects with stronger {\feiii} emission tend to have higher $L/L_{\rm Edd}$ ratios, consistent with the scenario outlined above. However, not all objects with high $L/L_{\rm Edd}$ display strong {\feiii} emission, suggesting that  $L/L_{\rm Edd}$ is not the sole driver of the {\feiii} equivalent width.

On the other hand, we also find that the equivalent width of UV34 anticorrelates with that of {\ion{He}{ii}} $\lambda$1640, which is another quantity known to show strong trends with {\civ} line properties.
{\ion{He}{ii}} is believed to trace the hardness of the SED which is illuminating the BLR, i.e. the strength of the extreme ultraviolet continuum radiation which is largely unseen by the observer. 
The observed correlation between {\feiii} and {\civ} could thus be driven by changes in the shape of the ionising SED.
However, recent work by
\citet{Ferland20}
has also pointed out that changes in the equivalent width of {\ion{He}{ii}} are also expected due to changes in the covering factor of the {\ion{He}{ii}} emitting gas, which could itself be varying in response to the hardness of the unseen SED, and thus the equivalent width of {\ion{He}{ii}} may not enjoy a linear relationship with the relative flux of Helium-ionising photons.

\section{Conclusions}
\label{sec:conclude}

While \feiii\ has long been known to be present in AGN spectra, 
the recent large set of atomic data  of
\citet{2014ApJ...785...99B} makes theoretical predictions of its emission
spectrum possible for the first time.  

Using \Cloudy, 
we have computed predictions for the strength
of the \feiii\ emission across a large range of parameter space, exploring the effect of variations in
the density, ionizing flux and turbulence of the emitting gas. 
The highly-excited  UV34 multiplet is  predicted to be the strongest feature, and the 
new predictions allow this multiplet to be accurately subtracted from the 
1909\,\AA\ $\ion{C}{iii}]+\ion{Fe}{iii}+\ion{Si}{iii}]$ complex. 
Our main results are as follows:
  
\begin{itemize}

 \item
 The lack of observed \feiii\ emission below 1000\,\AA\ shows that the \feiii\  lines in AGN are emitted by photoionized gas. 

  \item 
  Strong \feiii\ UV34 emission in quasar spectra demonstrates that high density $(n_H\simeq10^{14}\pcc)$ gas is present in the majority of luminous quasars.
  This gas must also produce large amounts of \ion{Al}{iii}\,$\lambda\lambda$1855,1863.
  
  \item 
  Together with the upper bounds on density from semi-forbidden lines such as \ciii], 
  our results show that the broad line region in AGN must have a non-uniform density structure.
 Our analysis suggests that the \feiii\ emission originates in gas about 1 dex closer to the black hole than the source of \civ \ emission.
  
  \item 
  Thermal line widths cannot produce sufficiently strong \feiii\ emission. Non-thermal
  line widths must therefore be present to increase the equivalent width by continuum pumping, i.e. increasing the number of
  continuum photons which can excite the transition.
  
  \item
  \citet{2004ApJ...615..610B} found that microturbulence also improves the fit
  to the \feii\ spectrum.
  \feiii\ is a more straightforward case since its spectrum is much simpler with
  isolated lines. 
  Together, the strength of the \feii\ and \feiii\ emission in quasar spectra make a compelling
  case for a turbulence in the line emitting region.
   
  \item Using the new atomic data, and our dense turbulent model, it is possible to predict the line ratios in the  \feiii\ UV34 multiplet:
  $\lambda  1895.46: \lambda 1914.06: \lambda 1926.30 = 1.316:1:0.921$.
  
  \item
  Using the observed strength of the  \ion{Al}{iii} doublet, 
  the \feiii\ UV34 multiplet can be modelled and subtracted from the 1909\,\AA\ complex 
  (equation~\ref{eq:ratios}).
  This enables a more accurate measurement of the \ion{C}{iii}] and 
  \ion{Si}{iii}] emission, even when these lines are  blended together.
  
\end{itemize}

\section*{Acknowledgements}


We thank the anonymous referee for a thoughtful report which led to improved clarity in many parts of the manuscript.
It is a pleasure to note useful comments from Bob Carswell, Gordon Richards, and Hagai Netzer.
MJT and ALR thank the Science and Technology Facilities Council (STFC) for the award of studentships.
GJF acknowledges support by NSF (1816537), NASA (ATP 17-ATP17-0141), and STScI (HST-AR-15018).
PCH acknowledges funding from STFC via the Institute of Astronomy, Cambridge, Consolidated Grant.
CPB acknowledges support through R1711APL : QUB Astronomy Observation and Theory Consolidated Grant.
NRB is funded by STFC Grant  ST/R000743/1 with the University of Strathclyde.
This work made use of Astropy \citep{astropy13, astropy18}, Matplotlib \citep{Hunter07}, and corner.py \citep{corner}.

Funding for the Sloan Digital Sky Survey IV has been provided by the Alfred P. Sloan Foundation, the U.S. Department of Energy Office of Science, and the Participating Institutions. SDSS-IV acknowledges
support and resources from the Center for High-Performance Computing at
the University of Utah. The SDSS web site is www.sdss.org.

SDSS-IV is managed by the Astrophysical Research Consortium for the 
Participating Institutions of the SDSS Collaboration including the 
Brazilian Participation Group, the Carnegie Institution for Science, 
Carnegie Mellon University, the Chilean Participation Group, the French Participation Group, Harvard-Smithsonian Center for Astrophysics, 
Instituto de Astrof\'isica de Canarias, The Johns Hopkins University, Kavli Institute for the Physics and Mathematics of the Universe (IPMU) / 
University of Tokyo, the Korean Participation Group, Lawrence Berkeley National Laboratory, 
Leibniz Institut f\"ur Astrophysik Potsdam (AIP),  
Max-Planck-Institut f\"ur Astronomie (MPIA Heidelberg), 
Max-Planck-Institut f\"ur Astrophysik (MPA Garching), 
Max-Planck-Institut f\"ur Extraterrestrische Physik (MPE), 
National Astronomical Observatories of China, New Mexico State University, 
New York University, University of Notre Dame, 
Observat\'ario Nacional / MCTI, The Ohio State University, 
Pennsylvania State University, Shanghai Astronomical Observatory, 
United Kingdom Participation Group,
Universidad Nacional Aut\'onoma de M\'exico, University of Arizona, 
University of Colorado Boulder, University of Oxford, University of Portsmouth, 
University of Utah, University of Virginia, University of Washington, University of Wisconsin, 
Vanderbilt University, and Yale University.




\bibliographystyle{mnras}
\bibliography{Paper_refs}

\begin{thebibliography}{}
\makeatletter
\relax
\def\mn@urlcharsother{\let\do\@makeother \do\$\do\&\do\#\do\^\do\_\do\%\do\~}
\def\mn@doi{\begingroup\mn@urlcharsother \@ifnextchar [ {\mn@doi@}
  {\mn@doi@[]}}
\def\mn@doi@[#1]#2{\def\@tempa{#1}\ifx\@tempa\@empty \href
  {http://dx.doi.org/#2} {doi:#2}\else \href {http://dx.doi.org/#2} {#1}\fi
  \endgroup}
\def\mn@eprint#1#2{\mn@eprint@#1:#2::\@nil}
\def\mn@eprint@arXiv#1{\href {http://arxiv.org/abs/#1} {{\tt arXiv:#1}}}
\def\mn@eprint@dblp#1{\href {http://dblp.uni-trier.de/rec/bibtex/#1.xml}
  {dblp:#1}}
\def\mn@eprint@#1:#2:#3:#4\@nil{\def\@tempa {#1}\def\@tempb {#2}\def\@tempc
  {#3}\ifx \@tempc \@empty \let \@tempc \@tempb \let \@tempb \@tempa \fi \ifx
  \@tempb \@empty \def\@tempb {arXiv}\fi \@ifundefined
  {mn@eprint@\@tempb}{\@tempb:\@tempc}{\expandafter \expandafter \csname
  mn@eprint@\@tempb\endcsname \expandafter{\@tempc}}}

\bibitem[\protect\citeauthoryear{{Astropy Collaboration} et~al.,}{{Astropy
  Collaboration} et~al.}{2013}]{astropy13}
{Astropy Collaboration} et~al., 2013, \mn@doi [\aap]
  {10.1051/0004-6361/201322068}, \href
  {http://adsabs.harvard.edu/abs/2013A%26A...558A..33A} {558, A33}

\bibitem[\protect\citeauthoryear{{Badnell} \& {Ballance}}{{Badnell} \&
  {Ballance}}{2014}]{2014ApJ...785...99B}
{Badnell} N.~R.,  {Ballance} C.~P.,  2014, \mn@doi [\apj]
  {10.1088/0004-637X/785/2/99}, \href
  {https://ui.adsabs.harvard.edu/abs/2014ApJ...785...99B} {785, 99}

\bibitem[\protect\citeauthoryear{{Baldwin}, {Ferland}, {Korista}  \&
  {Verner}}{{Baldwin} et~al.}{1995}]{1995ApJ...455L.119B}
{Baldwin} J.,  {Ferland} G.,  {Korista} K.,   {Verner} D.,  1995, \mn@doi
  [\apjl] {10.1086/309827}, \href
  {http://cdsads.u-strasbg.fr/abs/1995ApJ...455L.119B} {455, L119+}

\bibitem[\protect\citeauthoryear{{Baldwin} et~al.,}{{Baldwin}
  et~al.}{1996}]{1996ApJ...461..664B}
{Baldwin} J.~A.,  et~al., 1996, \mn@doi [\apj] {10.1086/177093}, \href
  {https://ui.adsabs.harvard.edu/abs/1996ApJ...461..664B} {461, 664}

\bibitem[\protect\citeauthoryear{{Baldwin}, {Ferland}, {Korista}, {Hamann}  \&
  {LaCluyz{\'e}}}{{Baldwin} et~al.}{2004}]{2004ApJ...615..610B}
{Baldwin} J.~A.,  {Ferland} G.~J.,  {Korista} K.~T.,  {Hamann} F.,
  {LaCluyz{\'e}} A.,  2004, \mn@doi [\apj] {10.1086/424683}, \href
  {http://cdsads.u-strasbg.fr/abs/2004ApJ...615..610B} {615, 610}

\bibitem[\protect\citeauthoryear{{Boroson} \& {Green}}{{Boroson} \&
  {Green}}{1992}]{BG92}
{Boroson} T.~A.,  {Green} R.~F.,  1992, \mn@doi [\apjs] {10.1086/191661}, \href
  {http://adsabs.harvard.edu/abs/1992ApJS...80..109B} {80, 109}

\bibitem[\protect\citeauthoryear{{Bottorff}, {Ferland}, {Baldwin}  \&
  {Korista}}{{Bottorff} et~al.}{2000}]{2000ApJ...542..644B}
{Bottorff} M.,  {Ferland} G.,  {Baldwin} J.,   {Korista} K.,  2000, \mn@doi
  [\apj] {10.1086/317051}, \href
  {http://cdsads.u-strasbg.fr/abs/2000ApJ...542..644B} {542, 644}

\bibitem[\protect\citeauthoryear{{Casebeer}, {Leighly}  \& {Baron}}{{Casebeer}
  et~al.}{2006}]{Casebeer06}
{Casebeer} D.~A.,  {Leighly} K.~M.,   {Baron} E.,  2006, \mn@doi [\apj]
  {10.1086/498125}, \href
  {https://ui.adsabs.harvard.edu/abs/2006ApJ...637..157C} {637, 157}

\bibitem[\protect\citeauthoryear{{Ferland}}{{Ferland}}{1992}]{1992ApJ...389L..63F}
{Ferland} G.~J.,  1992, \mn@doi [\apjl] {10.1086/186349}, \href
  {http://cdsads.u-strasbg.fr/abs/1992ApJ...389L..63F} {389, L63}

\bibitem[\protect\citeauthoryear{{Ferland} et~al.,}{{Ferland}
  et~al.}{2017}]{2017RMxAA..53..385F}
{Ferland} G.~J.,  et~al., 2017, \rmxaa, \href
  {https://ui.adsabs.harvard.edu/abs/2017RMxAA..53..385F} {53, 385}

\bibitem[\protect\citeauthoryear{{Ferland}, {Done}, {Jin}, {Landt}  \&
  {Ward}}{{Ferland} et~al.}{2020}]{Ferland20}
{Ferland} G.~J.,  {Done} C.,  {Jin} C.,  {Landt} H.,   {Ward} M.~J.,  2020,
  \mn@doi [\mnras] {10.1093/mnras/staa1207}, \href
  {https://ui.adsabs.harvard.edu/abs/2020MNRAS.tmp.1346F} {}

\bibitem[\protect\citeauthoryear{{Fian}, {Guerras}, {Mediavilla},
  {Jim{\'e}nez-Vicente}, {Mu{\~n}oz}, {Falco}, {Motta}  \& {Hanslmeier}}{{Fian}
  et~al.}{2018}]{2018ApJ...859...50F}
{Fian} C.,  {Guerras} E.,  {Mediavilla} E.,  {Jim{\'e}nez-Vicente} J.,
  {Mu{\~n}oz} J.~A.,  {Falco} E.~E.,  {Motta} V.,   {Hanslmeier} A.,  2018,
  \mn@doi [\apj] {10.3847/1538-4357/aabc0d}, \href
  {https://ui.adsabs.harvard.edu/abs/2018ApJ...859...50F} {859, 50}

\bibitem[\protect\citeauthoryear{Foreman-Mackey}{Foreman-Mackey}{2016}]{corner}
Foreman-Mackey D.,  2016, \mn@doi [The Journal of Open Source Software]
  {10.21105/joss.00024}, 24

\bibitem[\protect\citeauthoryear{{Francis}, {Hewett}, {Foltz}, {Chaffee},
  {Weymann}  \& {Morris}}{{Francis} et~al.}{1991}]{Francis91}
{Francis} P.~J.,  {Hewett} P.~C.,  {Foltz} C.~B.,  {Chaffee} F.~H.,  {Weymann}
  R.~J.,   {Morris} S.~L.,  1991, \mn@doi [\apj] {10.1086/170066}, \href
  {https://ui.adsabs.harvard.edu/abs/1991ApJ...373..465F} {373, 465}

\bibitem[\protect\citeauthoryear{{Giustini} \& {Proga}}{{Giustini} \&
  {Proga}}{2019}]{2019A&A...630A..94G}
{Giustini} M.,  {Proga} D.,  2019, \mn@doi [\aap]
  {10.1051/0004-6361/201833810}, \href
  {https://ui.adsabs.harvard.edu/abs/2019A&A...630A..94G} {630, A94}

\bibitem[\protect\citeauthoryear{{Graham}, {Clowes}  \& {Campusano}}{{Graham}
  et~al.}{1996}]{Graham96}
{Graham} M.~J.,  {Clowes} R.~G.,   {Campusano} L.~E.,  1996, \mn@doi [\mnras]
  {10.1093/mnras/279.4.1349}, \href
  {https://ui.adsabs.harvard.edu/abs/1996MNRAS.279.1349G} {279, 1349}

\bibitem[\protect\citeauthoryear{{Grier} et~al.,}{{Grier}
  et~al.}{2019}]{Grier19}
{Grier} C.~J.,  et~al., 2019, \mn@doi [\apj] {10.3847/1538-4357/ab4ea5}, \href
  {https://ui.adsabs.harvard.edu/abs/2019ApJ...887...38G} {887, 38}

\bibitem[\protect\citeauthoryear{{Hu} et~al.,}{{Hu} et~al.}{2015}]{Hu15}
{Hu} C.,  et~al., 2015, \mn@doi [\apj] {10.1088/0004-637X/804/2/138}, \href
  {https://ui.adsabs.harvard.edu/abs/2015ApJ...804..138H} {804, 138}

\bibitem[\protect\citeauthoryear{Hunter}{Hunter}{2007}]{Hunter07}
Hunter J.~D.,  2007, \mn@doi [Computing In Science \& Engineering]
  {10.1109/MCSE.2007.55}, 9, 90

\bibitem[\protect\citeauthoryear{{Jin}, {Ward}  \& {Done}}{{Jin}
  et~al.}{2012}]{2012MNRAS.425..907J}
{Jin} C.,  {Ward} M.,   {Done} C.,  2012, \mn@doi [\mnras]
  {10.1111/j.1365-2966.2012.21272.x}, \href
  {https://ui.adsabs.harvard.edu/abs/2012MNRAS.425..907J} {425, 907}

\bibitem[\protect\citeauthoryear{{Korista}, {Baldwin}, {Ferland}  \&
  {Verner}}{{Korista} et~al.}{1997}]{1997ApJS..108..401K}
{Korista} K.,  {Baldwin} J.,  {Ferland} G.,   {Verner} D.,  1997, \mn@doi
  [\apjs] {10.1086/312966}, \href
  {http://cdsads.u-strasbg.fr/abs/1997ApJS..108..401K} {108, 401}

\bibitem[\protect\citeauthoryear{Kramida, {Yu.~Ralchenko}, Reader  \& {and NIST
  ASD Team}}{Kramida et~al.}{2018}]{NIST_ASD}
Kramida A.,  {Yu.~Ralchenko} Reader J.,   {and NIST ASD Team} 2018, {NIST
  Atomic Spectra Database (ver. 5.6.1), [Online]. Available:
  {\tt{https://physics.nist.gov/asd}} [2019, September 6]. National Institute
  of Standards and Technology, Gaithersburg, MD.}

\bibitem[\protect\citeauthoryear{{Laha}, {Tyndall}, {Keenan}, {Ballance},
  {Ramsbottom}, {Ferland }  \& {Hibbert}}{{Laha} et~al.}{2017}]{Laha17}
{Laha} S.,  {Tyndall} N.~B.,  {Keenan} F.~P.,  {Ballance} C.~P.,  {Ramsbottom}
  C.~A.,  {Ferland } G.~J.,   {Hibbert} A.,  2017, \mn@doi [\apj]
  {10.3847/1538-4357/aa7071}, \href
  {https://ui.adsabs.harvard.edu/abs/2017ApJ...841....3L} {841, 3}

\bibitem[\protect\citeauthoryear{{Lykins}, {Ferland}, {Porter}, {van Hoof},
  {Williams}  \& {Gnat}}{{Lykins}
  et~al.}{2013}]{Lykins.M13Radiative-cooling-in-collisionally-ionized}
{Lykins} M.~L.,  {Ferland} G.~J.,  {Porter} R.~L.,  {van Hoof} P.~A.~M.,
  {Williams} R.~J.~R.,   {Gnat} O.,  2013, \mnras, 429, 3133

\bibitem[\protect\citeauthoryear{{Lykins} et~al.,}{{Lykins}
  et~al.}{2015}]{2015ApJ...807..118L}
{Lykins} M.~L.,  et~al., 2015, \mn@doi [\apj] {10.1088/0004-637X/807/2/118},
  \href {https://ui.adsabs.harvard.edu/abs/2015ApJ...807..118L} {807, 118}

\bibitem[\protect\citeauthoryear{{Marziani} \& {Sulentic}}{{Marziani} \&
  {Sulentic}}{2014}]{2014MNRAS.442.1211M}
{Marziani} P.,  {Sulentic} J.~W.,  2014, \mn@doi [\mnras]
  {10.1093/mnras/stu951}, \href
  {https://ui.adsabs.harvard.edu/abs/2014MNRAS.442.1211M} {442, 1211}

\bibitem[\protect\citeauthoryear{{Mediavilla}, {Jim{\'e}nez-Vicente}, {Fian},
  {Mu{\~n}oz}, {Falco}, {Motta}  \& {Guerras}}{{Mediavilla}
  et~al.}{2018}]{2018ApJ...862..104M}
{Mediavilla} E.,  {Jim{\'e}nez-Vicente} J.,  {Fian} C.,  {Mu{\~n}oz} J.~A.,
  {Falco} E.,  {Motta} V.,   {Guerras} E.,  2018, \mn@doi [\apj]
  {10.3847/1538-4357/aacbd3}, \href
  {https://ui.adsabs.harvard.edu/abs/2018ApJ...862..104M} {862, 104}

\bibitem[\protect\citeauthoryear{{Mediavilla}, {Jim{\'e}nez-Vicente},
  {Mej{\'\i}a-Restrepo}, {Motta}, {Falco}, {Mu{\~n}oz}, {Fian}  \&
  {Guerras}}{{Mediavilla} et~al.}{2019}]{2019ApJ...880...96M}
{Mediavilla} E.,  {Jim{\'e}nez-Vicente} J.,  {Mej{\'\i}a-Restrepo} J.,  {Motta}
  V.,  {Falco} E.,  {Mu{\~n}oz} J.~A.,  {Fian} C.,   {Guerras} E.,  2019,
  \mn@doi [\apj] {10.3847/1538-4357/ab2910}, \href
  {https://ui.adsabs.harvard.edu/abs/2019ApJ...880...96M} {880, 96}

\bibitem[\protect\citeauthoryear{{Moore}}{{Moore}}{1952}]{Moore52}
{Moore} C.~E.,  1952, {An ultraviolet multiplet table}

\bibitem[\protect\citeauthoryear{{Nikoli{\'c}} et~al.,}{{Nikoli{\'c}}
  et~al.}{2018}]{2018ApJS..237...41N}
{Nikoli{\'c}} D.,  et~al., 2018, \mn@doi [\apjs] {10.3847/1538-4365/aad3c5},
  \href {https://ui.adsabs.harvard.edu/abs/2018ApJS..237...41N} {237, 41}

\bibitem[\protect\citeauthoryear{{Osterbrock} \& {Ferland}}{{Osterbrock} \&
  {Ferland}}{2006}]{2006agna.book.....O}
{Osterbrock} D.~E.,  {Ferland} G.~J.,  2006, {Astrophysics of gaseous nebulae
  and active galactic nuclei}

\bibitem[\protect\citeauthoryear{{P{\^a}ris} et~al.,}{{P{\^a}ris}
  et~al.}{2017}]{2017A&A...597A..79P}
{P{\^a}ris} I.,  et~al., 2017, \mn@doi [\aap] {10.1051/0004-6361/201527999},
  \href {https://ui.adsabs.harvard.edu/abs/2017A&A...597A..79P} {597, A79}

\bibitem[\protect\citeauthoryear{{P{\^a}ris} et~al.,}{{P{\^a}ris}
  et~al.}{2018}]{DR14Q}
{P{\^a}ris} I.,  et~al., 2018, \mn@doi [\aap] {10.1051/0004-6361/201732445},
  \href {https://ui.adsabs.harvard.edu/abs/2018A&A...613A..51P} {613, A51}

\bibitem[\protect\citeauthoryear{{Price-Whelan} et~al.,}{{Price-Whelan}
  et~al.}{2018}]{astropy18}
{Price-Whelan} A.~M.,  et~al., 2018, \mn@doi [\aj] {10.3847/1538-3881/aabc4f},
  \href {https://ui.adsabs.harvard.edu/#abs/2018AJ....156..123T} {156, 123}

\bibitem[\protect\citeauthoryear{{Rankine}, {Hewett}, {Banerji}  \&
  {Richards}}{{Rankine} et~al.}{2020}]{Rankine20}
{Rankine} A.~L.,  {Hewett} P.~C.,  {Banerji} M.,   {Richards} G.~T.,  2020,
  \mn@doi [\mnras] {10.1093/mnras/staa130}, \href
  {https://ui.adsabs.harvard.edu/abs/2020MNRAS.492.4553R} {492, 4553}

\bibitem[\protect\citeauthoryear{{Richards} et~al.,}{{Richards}
  et~al.}{2011}]{Richards11}
{Richards} G.~T.,  et~al., 2011, \mn@doi [\aj] {10.1088/0004-6256/141/5/167},
  \href {https://ui.adsabs.harvard.edu/abs/2011AJ....141..167R} {141, 167}

\bibitem[\protect\citeauthoryear{{Shen} et~al.,}{{Shen} et~al.}{2011}]{Shen11}
{Shen} Y.,  et~al., 2011, \mn@doi [\apjs] {10.1088/0067-0049/194/2/45}, \href
  {http://adsabs.harvard.edu/abs/2011ApJS..194...45S} {194, 45}

\bibitem[\protect\citeauthoryear{{Stevans}, {Shull}, {Danforth}  \&
  {Tilton}}{{Stevans} et~al.}{2014}]{2014ApJ...794...75S}
{Stevans} M.~L.,  {Shull} J.~M.,  {Danforth} C.~W.,   {Tilton} E.~M.,  2014,
  \mn@doi [\apj] {10.1088/0004-637X/794/1/75}, \href
  {https://ui.adsabs.harvard.edu/abs/2014ApJ...794...75S} {794, 75}

\bibitem[\protect\citeauthoryear{{Vanden Berk} et~al.,}{{Vanden Berk}
  et~al.}{2001}]{VandenBerk01}
{Vanden Berk} D.~E.,  et~al., 2001, \mn@doi [\aj] {10.1086/321167}, \href
  {https://ui.adsabs.harvard.edu/abs/2001AJ....122..549V} {122, 549}

\bibitem[\protect\citeauthoryear{{Vestergaard} \& {Osmer}}{{Vestergaard} \&
  {Osmer}}{2009}]{Vestergaard09}
{Vestergaard} M.,  {Osmer} P.~S.,  2009, \mn@doi [\apj]
  {10.1088/0004-637X/699/1/800}, \href
  {https://ui.adsabs.harvard.edu/abs/2009ApJ...699..800V} {699, 800}

\bibitem[\protect\citeauthoryear{{Vestergaard} \& {Wilkes}}{{Vestergaard} \&
  {Wilkes}}{2001}]{2001ApJS..134....1V}
{Vestergaard} M.,  {Wilkes} B.~J.,  2001, \mn@doi [\apjs] {10.1086/320357},
  \href {https://ui.adsabs.harvard.edu/abs/2001ApJS..134....1V} {134, 1}

\bibitem[\protect\citeauthoryear{{Wills}, {Netzer}  \& {Wills}}{{Wills}
  et~al.}{1985}]{Wills1985}
{Wills} B.~J.,  {Netzer} H.,   {Wills} D.,  1985, \mn@doi [\apj]
  {10.1086/162767}, \href {http://cdsads.u-strasbg.fr/abs/1985ApJ...288...94W}
  {288, 94}

\bibitem[\protect\citeauthoryear{{Zhang} et~al.,}{{Zhang}
  et~al.}{2019}]{Zhang19}
{Zhang} Z.-X.,  et~al., 2019, \mn@doi [\apj] {10.3847/1538-4357/ab1099}, \href
  {https://ui.adsabs.harvard.edu/abs/2019ApJ...876...49Z} {876, 49}

\makeatother
\end{thebibliography}




\appendix
\section{Atomic data sources}

Version 17 of Cloudy obtains most of its atomic data from external data files
\citep{2017RMxAA..53..385F}.  
Our ``Stout'' data format is described in \citet{2015ApJ...807..118L}.
This structure makes it simple to ``drop in'' large data sets such as the \feiii\
data presented in \citet{2014ApJ...785...99B}.
Those data were published in ADF04 format, a common structure used
for atomic data exchange, and the Cloudy project has scripts to convert
ADF04 to our Stout format.

High densities are considered in this investigation.  
\citet{2017RMxAA..53..385F} describes our ``equivalent two-level atom'' model
used to derive the ionization of many-electron systems.  
This works in terms of total summed recombination rate coefficients.
Recombination processes are suppressed at high densities and our approach
in treating this process is described in \citet{2018ApJS..237...41N}.
This physics is highly uncertain and we have experimented by changing
the suppression factor by $\pm 0.5$~dex.
The conclusions reached in this paper were unchanged.

\section{Systemic Redshift Estimation}

The systemic redshifts used in this work are those described by \citet{Rankine20}, which do not make use of the \civ\ line. 
This approach avoids biasing the systemic redshift estimation in cases where the \civ\ emission profile is skewed, which we have found to be an issue with the SDSS pipeline redshifts.
Using our redshifts, and the \feiii\ energy level data now included in \Cloudy, we find that the \feiii\ emission in our sample is consistent with coming from gas at the systemic redshift. In particular, we are unable to reproduce the result of \citet{2018ApJ...862..104M}, who find that the \feiii\ $\lambda 2075$ emission lines are systematically redshifted, and the magnitude of this additional redshift is correlated with the full width at half maximum (FWHM) of the \civ\ emission line. 

\begin{figure}
\begin{center}
\includegraphics[clip=on,width=\columnwidth,
keepaspectratio]{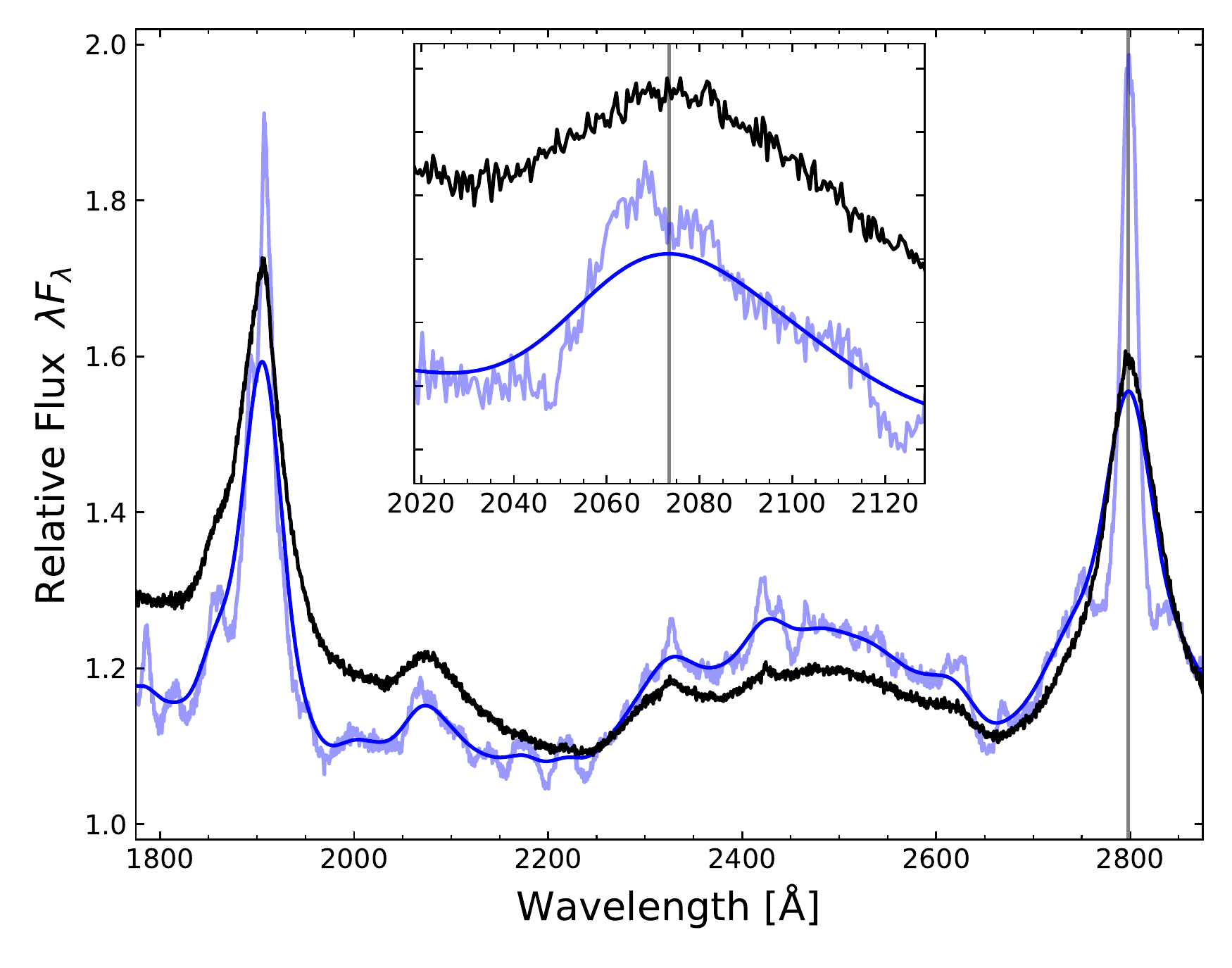}
\end{center}
\caption{Composite of objects in our sample with FWHM(\mgii) $\simeq$3800\kmps (blue) and $\simeq$10\,200\kmps (black).
The narrow-lined composite has been broadened with a Gaussian kernel to match the velocity widths.
Inset: the location of peak of the \feiii\ complex is seen at at 2073.5\,\AA\ both composites.}
\label{fig:Composite2}
\end{figure}

In Fig.~\ref{fig:Composite2}, we show composite spectra constructed from those objects in our sample which are also contained within the twelfth data release of the SDSS, of which the quasar catalogue \citep{2017A&A...597A..79P} reports the FWHM of the \mgii\ emission line.
We construct two composites: one using  375 objects with FWHM(\mgii) in the range 2000-4000\,\kms, and one using 967 objects with FWHM(\mgii) in the range 9000-15\,000\,\kms. The median FWHM(\mgii) in each subsample is 3800 and 10\,200\kmps respectively. The narrow-lined composite is artificially broadened using a Gaussian kernel to match the velocity width of the broader-lined subsample. 
The work of \citet[][their Fig. 2]{2019ApJ...880...96M} would suggest that the factor of $\simeq$2.7 difference in FWHM(\mgii) would correspond to a shift of at least 10\,\AA\ in the location of the \feiii\ lines between the two composites, which we would expect to be easily detectable.
We see that the peak of \mgii\ is observed at 2798\,\AA\ in both our composites, and there is no noticable difference between the peak of the \feiii\ complex  marked at 2073.5\,\AA.

\section{Modelling the 1909\,\AA\ complex}
\label{sec:example_fits}

\begin{figure*}
\begin{center}
\includegraphics[trim=0 10 10 55,clip,width=\columnwidth,keepaspectratio]{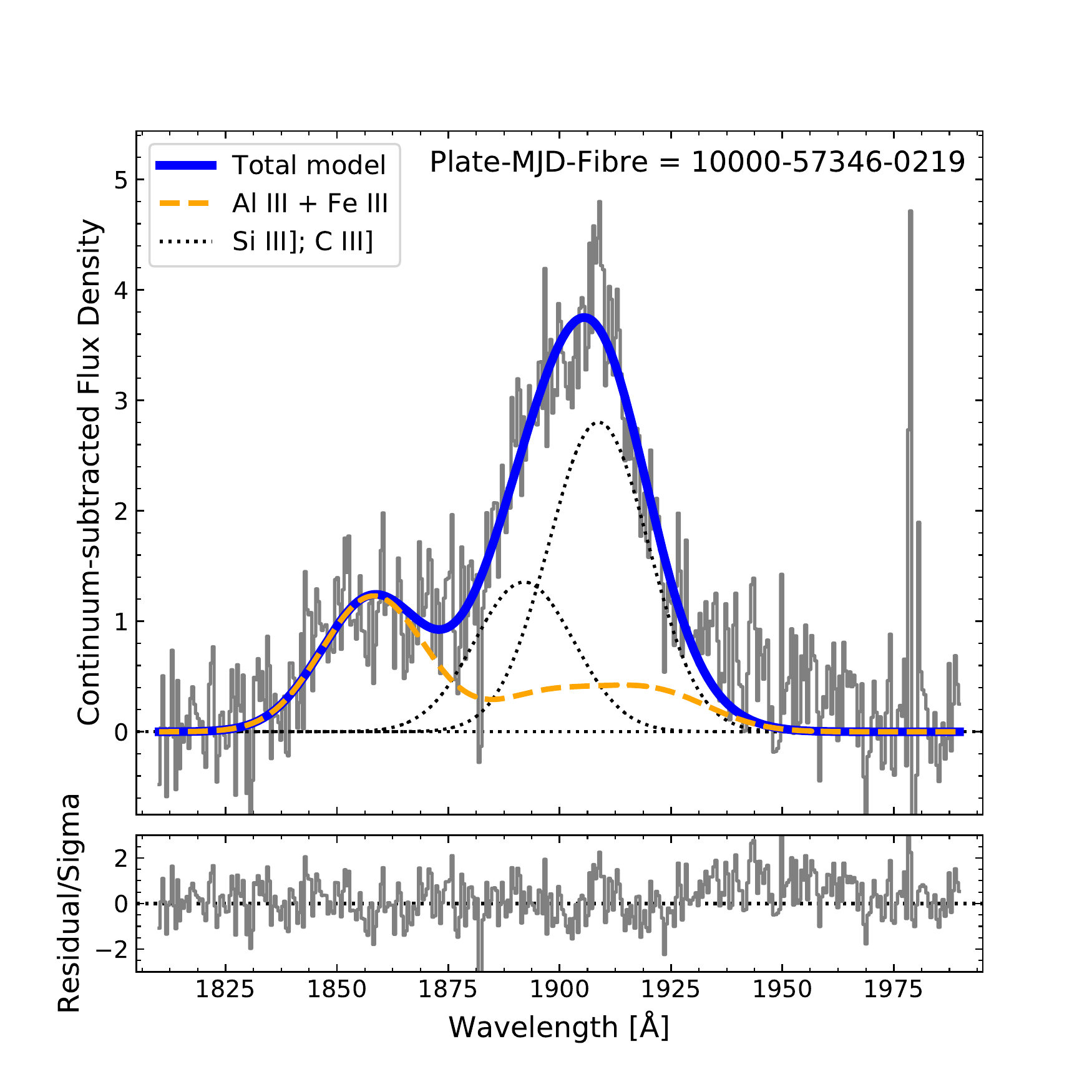}
\includegraphics[trim=0 10 10 55,clip,width=\columnwidth,keepaspectratio]{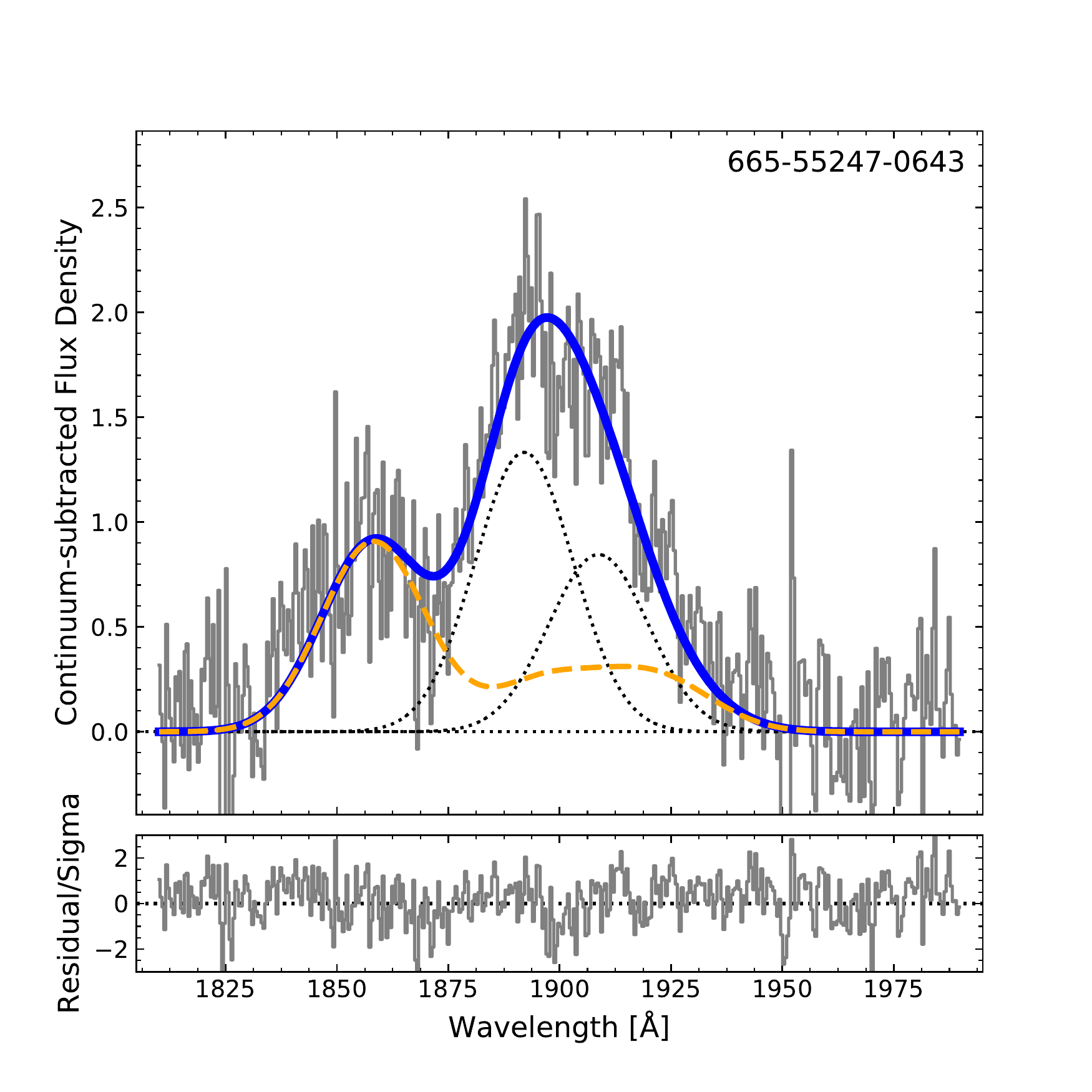}
\includegraphics[trim=0 10 10 55,clip,width=\columnwidth,keepaspectratio]{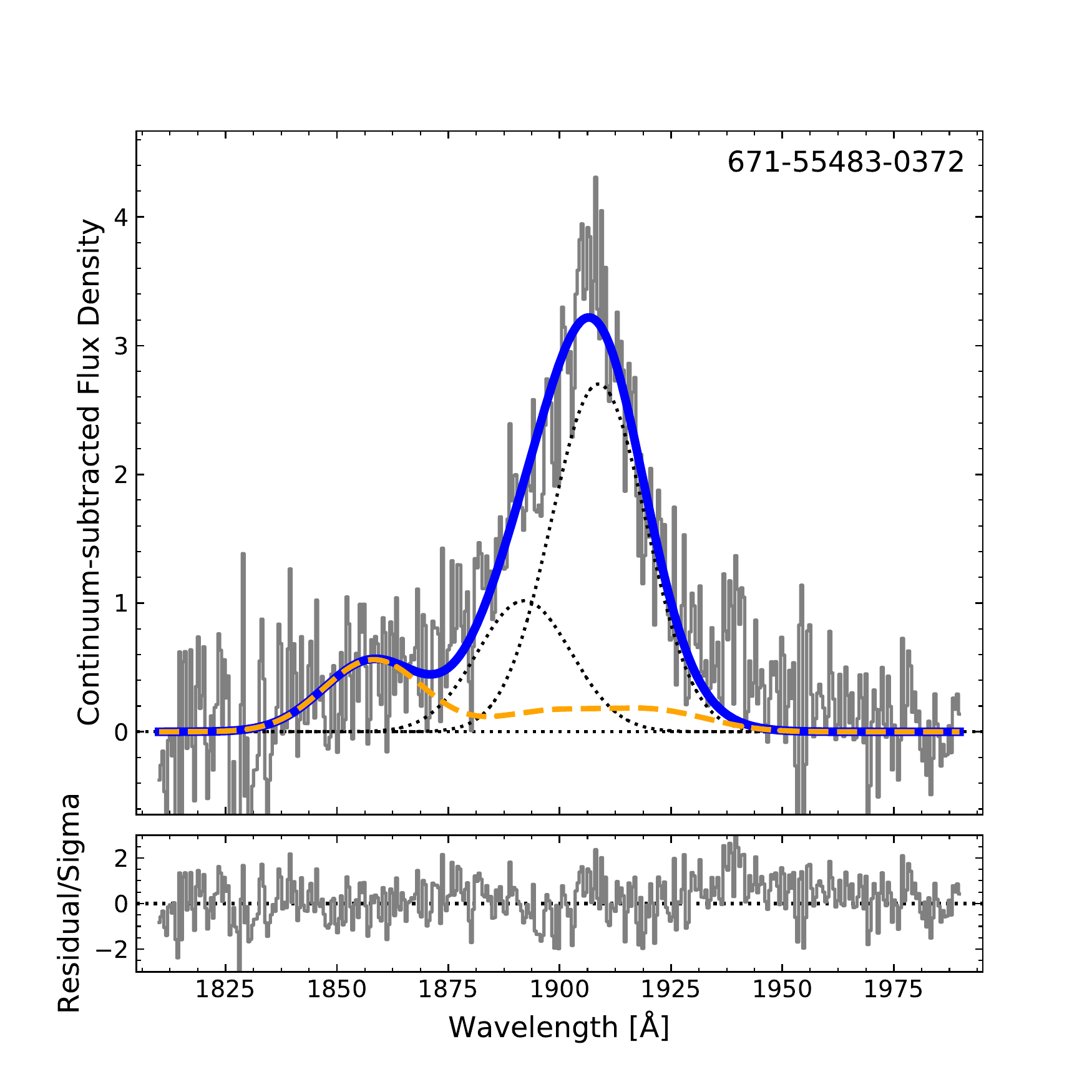}
\includegraphics[trim=0 10 10 55,clip,width=\columnwidth,keepaspectratio]{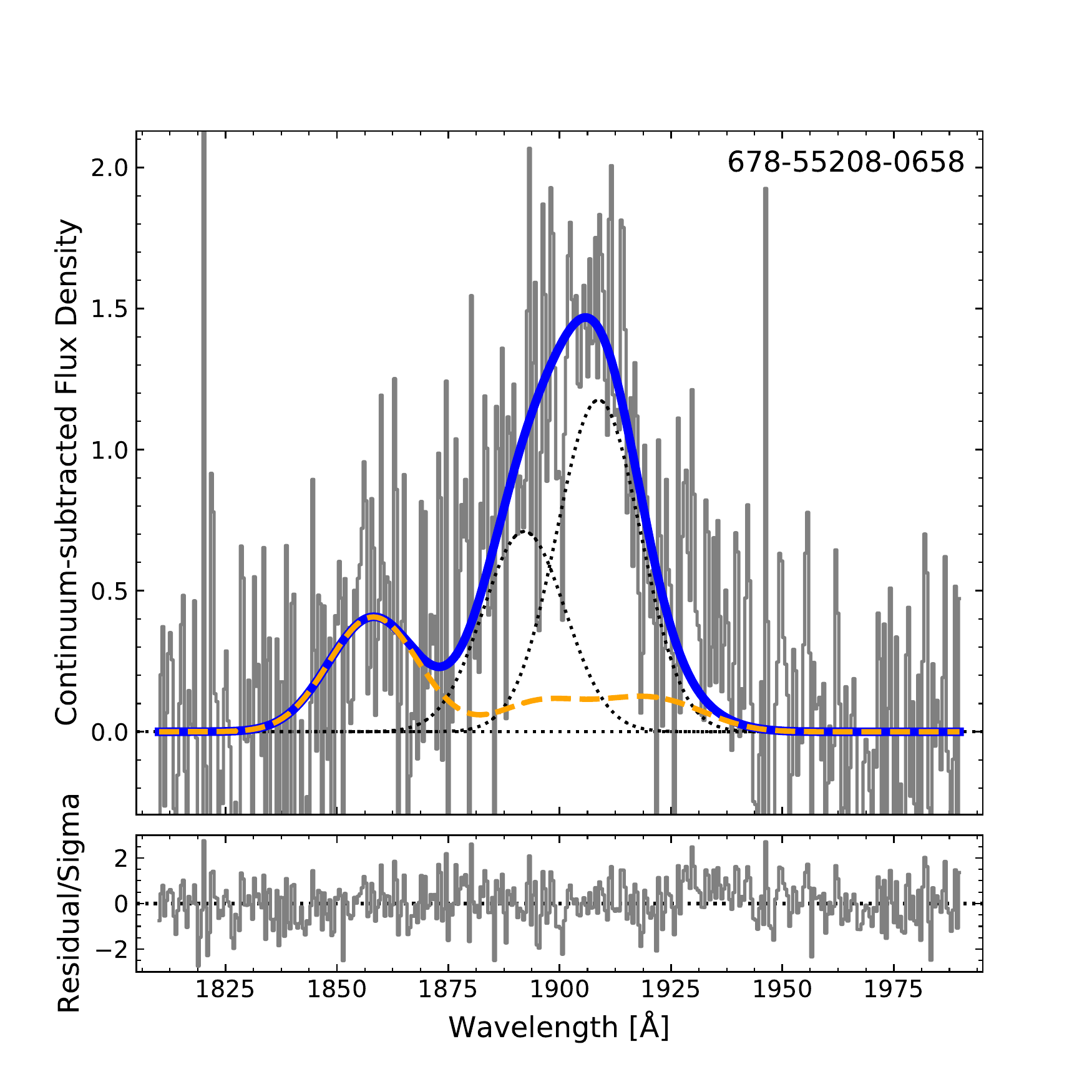}
\includegraphics[trim=0 20 10 55,clip,width=\columnwidth,keepaspectratio]{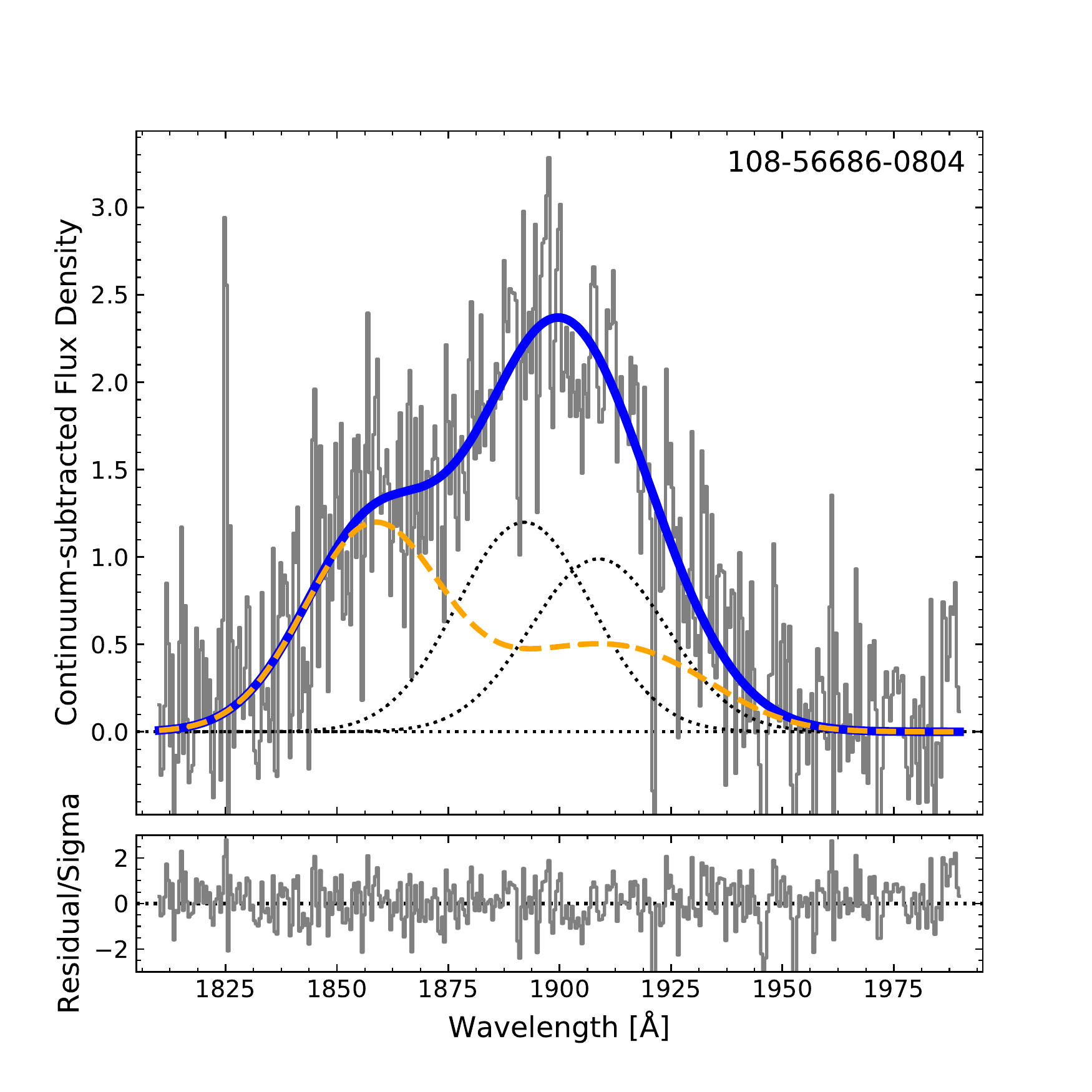}
\includegraphics[trim=0 20 10 55,clip,width=\columnwidth,keepaspectratio]{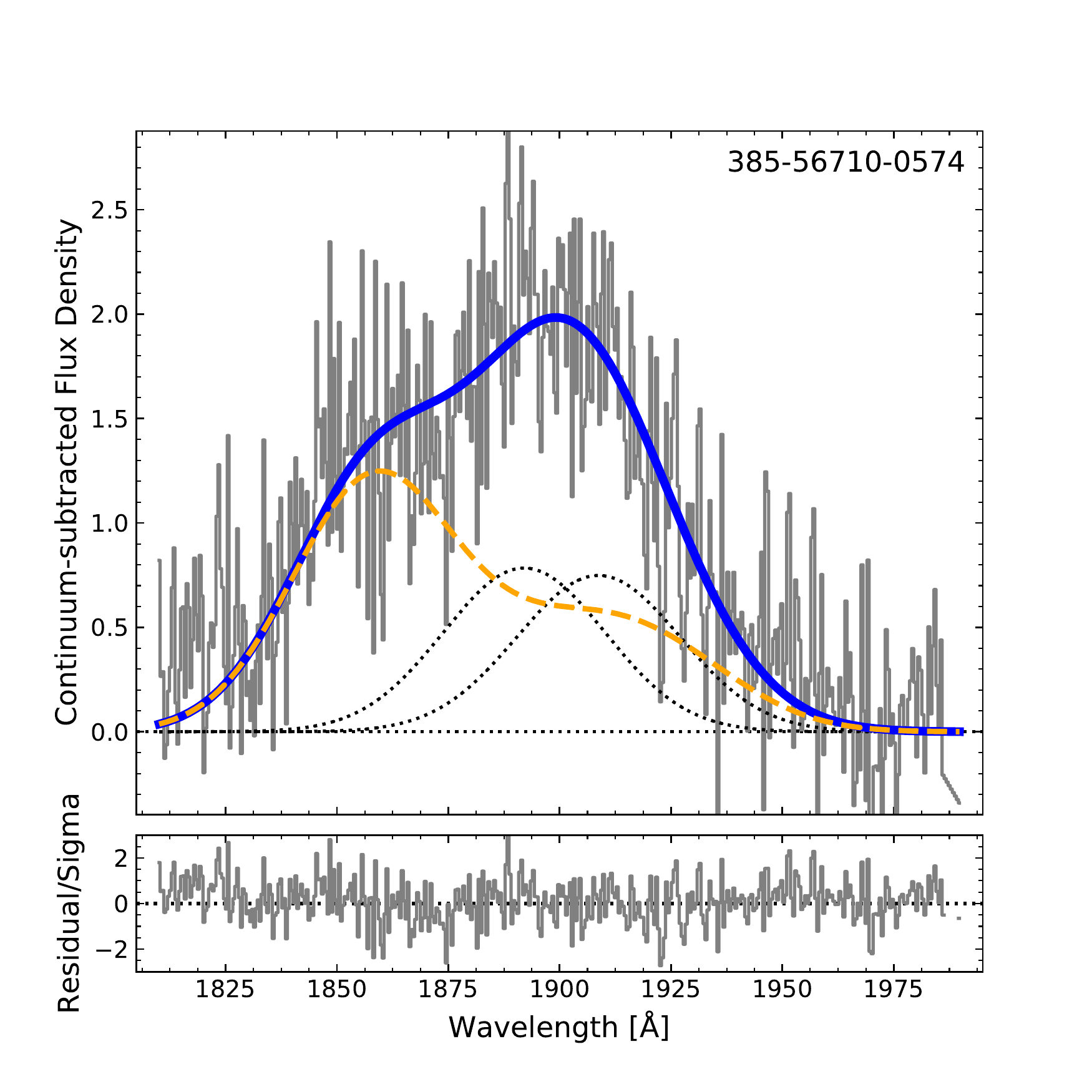}
\end{center}
\caption{Examples of our best fitting models to the 1909\,\AA\ complex. The \ciii] emission may be slightly narrower than the \ion{Al}{iii} and \feiii\ emission in e.g. 671-55483-0372, but in general the S/N is such that constraining the velocity widths of all lines to be equal provides a reasonable fit to the data.}
\label{fig:example_fits}
\end{figure*}

In Fig.~\ref{fig:example_fits},
we present six examples of our best fitting models for the 1909\,{\AA} emission complex, as described in Section~\ref{sec:line_fits}.
Each model is fit to the data in the wavelength range 1820-1920\,{\AA}, avoiding contamination from weak {\feii} emission features at $\simeq$1940\,\AA.

\bsp    
\label{lastpage}
\end{document}